\documentclass[pdflatex,sn-mathphys-num]{sn-jnl}

\usepackage{fix-cm}

\usepackage{hyperref}
\usepackage{float}
\usepackage{verbatim} 
\usepackage{apalike}
\usepackage{color}
\usepackage{tabularx}
\usepackage{booktabs}
\usepackage{threeparttable}
\usepackage{makecell}
\usepackage{longtable}
\usepackage{array}

\usepackage{amsmath}
\usepackage{bbm}

\usepackage{multirow}
\usepackage{graphicx}
\usepackage{lineno}
\usepackage{graphicx}
\usepackage{multirow}
\usepackage{dblfloatfix} 
\usepackage[dvipsnames]{xcolor}
\usepackage{soul}
\usepackage{color}
\usepackage{url}
\usepackage{float}
\usepackage{boldline}
\usepackage{xcolor,colortbl}
\usepackage{tikz}

\usepackage{xspace}

\usepackage{algorithm}
\usepackage{algpseudocode}

\usepackage{graphicx}
\usepackage{multirow}
\usepackage{amsmath,amssymb,amsfonts}
\usepackage{amsthm}
\usepackage{mathrsfs}
\usepackage[title]{appendix}
\usepackage[dvipsnames,table]{xcolor}
\usepackage{textcomp}
\usepackage{manyfoot}
\usepackage{booktabs}
\usepackage{algorithm}
\usepackage{algorithmicx}
\usepackage{algpseudocode}
\usepackage{listings}

\usepackage{tikz}
\usepackage{pgfplots}
\usepackage{pgfplots}
\usepackage{pgfplots}
\pgfplotsset{compat=1.18}
\usepackage{lineno}
\usepackage{dblfloatfix}
\usepackage{soul}
\usepackage{url}
\usepackage{float}
\usepackage{boldline}
\usepackage{pifont}

\theoremstyle{thmstyleone}%
\theoremstyle{thmstyletwo}%

\theoremstyle{thmstylethree}%

\usepackage{bbm}
\usepackage{makecell}
\usepackage{tabularx}
\usepackage{booktabs}
\usepackage{array}

\setcitestyle{authoryear,round,semicolon,aysep={,}}
\hypersetup{
    colorlinks=true,
    citecolor=blue
}

\begin{document}

\title[Article Title]{Reliability-Aware and Anatomy-Preserving Replay for Continual Echocardiographic Segmentation}


\author[1]{\fnm{Zahid} \sur{Ullah}}\email{zahid1989@dongguk.edu}

\author[1]{\fnm{Sieun} \sur{Choi}}\email{sieunchoi@dgu.ac.kr}

\author*[1]{\fnm{Jihie} \sur{Kim}}\email{jihie.kim@dgu.edu}

\affil*[1]{\orgdiv{Department of Computer Science and Artificial Intelligence}, \orgname{Dongguk University}, 
\city{Seoul}, \postcode{04620}, \country{South Korea}}




\abstract{
Continual echocardiographic segmentation remains challenging due to catastrophic forgetting, domain shifts, and heterogeneous anatomical supervision. Existing continual-learning methods often struggle to preserve previously acquired knowledge while adapting to new imaging distributions, particularly under limited replay memory and incomplete anatomical annotations. To address these challenges, we propose the Reliability-Aware Replay Network (RAR-Net), a continual segmentation framework integrating multi-factor exemplar prioritization, protected task-aware memory, forgetting-conditioned adaptive replay, and structure-aware supervision. RAR-Net combines class, boundary, image, anatomical, and temporal reliability criteria to prioritize informative historical examples, while adaptive replay regulates their contribution during sequential training. Structure-aware supervision further reinforces previously learned anatomical knowledge when subsequent datasets provide incomplete annotations. Extensive experiments performed on CAMUS, CardiacNet, and EchoNet-Dynamic demonstrate that our RAR-Net achieves the highest final average accuracy among the evaluated continual-learning methods, with competitive historical knowledge retention and segmentation performance approaching the offline joint-training reference. These findings indicate that coordinating reliability-guided memory management, forgetting-conditioned replay, and anatomy-specific supervision can support knowledge retention and adaptation under the evaluated sequence of heterogeneous echocardiographic datasets.
}




\maketitle

\section{Introduction}
\label{sec:introduction}

\begin{figure}[!t]
    \centering
    \includegraphics[width=0.5\columnwidth]{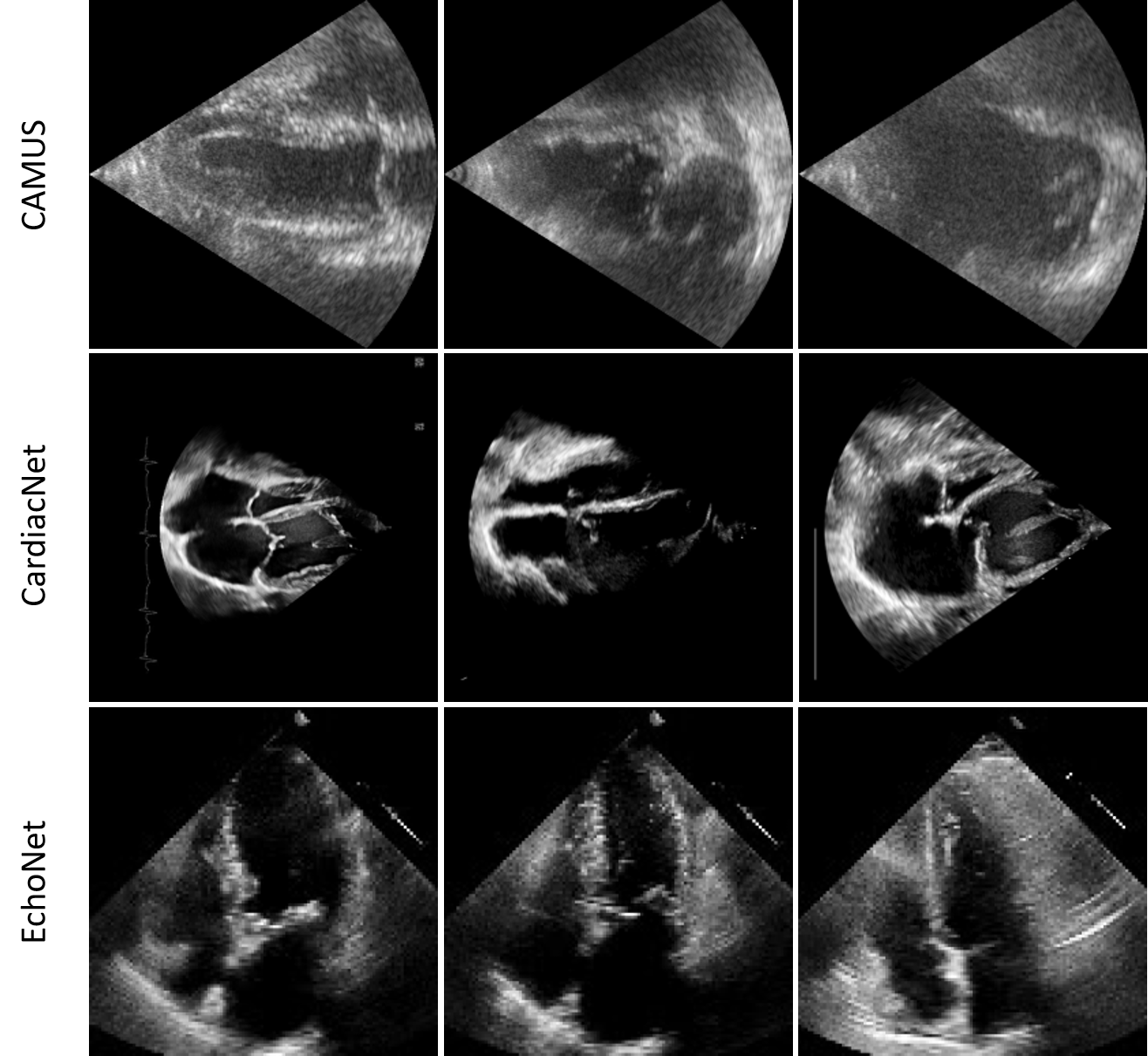}
    \caption{Representative echocardiographic images from CAMUS, CardiacNet, and EchoNet-Dynamic, shown in the top, middle, and bottom rows, respectively, illustrating differences in appearance, field of view, and anatomical presentation.}
    \label{Fig1}
\end{figure}

Cardiovascular diseases remain a leading cause of mortality worldwide \cite{da2022cascade,du2026rspc}. Echocardiography is widely used for cardiovascular assessment because it is non-invasive, accessible, cost-effective, and capable of visualizing cardiac anatomy and motion in real time. However, automated echocardiographic analysis remains challenging due to speckle noise, acoustic shadowing, low tissue contrast, ambiguous boundaries, and variability across patients, scanners, institutions, and acquisition protocols \cite{leclerc2019deep,ouyang2020video,deng2024memsam}. Accurate segmentation of the left ventricle (LV), myocardium (MYO), and left atrium (LA) supports quantitative assessment of cardiac morphology and function, including ventricular volume, ejection fraction, and structural remodeling. Figure~\ref{Fig1} illustrates the substantial appearance and anatomical variability across CAMUS, CardiacNet, and EchoNet-Dynamic. These differences reflect the domain shifts encountered by segmentation systems operating across evolving clinical environments, where maintaining previously acquired anatomical knowledge is as important as adapting to new data.

Deep learning has substantially advanced cardiac segmentation \cite{bernard2018deep,zhuang2019evaluation}. Encoder-decoder architectures, particularly U-Net and its variants \cite{ronneberger2015u,wang2025alda}, provide effective spatial representations, while CAMUS and EchoNet-Dynamic have established important benchmarks for multi-structure segmentation and echocardiographic video analysis \cite{leclerc2019deep,ouyang2020video}. Attention- and Transformer-based architectures further improve long-range feature modeling \cite{chen2024transunet,xiao2026pia,khan2026giaformer}, and medical foundation models such as MedSAM demonstrate segmentation capabilities across heterogeneous imaging settings \cite{ma2024segment}. Temporal modeling has also improved echocardiographic analysis by exploiting relationships between cardiac frames \cite{deng2024memsam,lin2024dynamic}. Nevertheless, most approaches assume a relatively static training distribution. In practice, changes in patient populations, imaging devices, acquisition protocols, and annotation standards require repeated model adaptation. Retraining on all accumulated data may be computationally expensive or restricted by data storage, governance, and privacy constraints \cite{li2022domain,wu2024continual,kumari2025continual}. 

Continual learning (CL) addresses this challenge by updating models sequentially while retaining knowledge from previously encountered tasks \cite{parisi2019continual,de2021continual}. Its central difficulty is balancing plasticity for new distributions with stability of previously learned representations, thereby limiting catastrophic forgetting \cite{kirkpatrick2017overcoming,li2017learning}. This challenge becomes particularly pronounced in medical segmentation because domain shifts, restricted historical data, and heterogeneous annotations can simultaneously affect spatial and anatomical knowledge \cite{wu2024continual,kumari2025continual}.
Although these approaches can reduce forgetting, parameter- or prediction-level constraints do not necessarily preserve the anatomical details required for dense segmentation. Replay-based methods retain historical examples or reconstruct previous distributions during subsequent training \cite{rebuffi2017icarl,perkonigg2021dynamic,li2022domain,ranem2024atlas,Joham2025FOCUS}. 
Despite these advances, effective replay depends not only on retaining historical information but also on selecting informative samples, maintaining domain representation, and controlling their influence during subsequent learning.

A major challenge is identifying which examples should occupy a bounded replay memory. Random selection may retain redundant samples, while single-criterion selection can overlook complementary anatomical information. In echocardiography, replay value depends on several factors: the presence of relevant cardiac structures, boundary information \cite{zhao2024boundary,lin2024boundary}, image quality \cite{nazar2024machine,van2025regional}, anatomical plausibility \cite{yeganeh2023scope,wyburd2024anatomically}, and temporal relationships \cite{deng2024memsam,lin2024dynamic}. These observations motivate the five-factor reliability representation adopted in the proposed Reliability-Aware Replay Network (RAR-Net), combining class, boundary, image, anatomical, and temporal evidence to prioritize historical examples.

However, reliability-based selection alone cannot guarantee balanced memory representation. Larger or recently observed datasets may dominate a shared buffer, reducing coverage of earlier domains. Building on the importance of diversity-aware memory \cite{perkonigg2021dynamic}, RAR-Net introduces protected task-aware memory partitions that maintain historical-domain representation while prioritizing informative samples within each task. Another limitation is the fixed contribution assigned to historical replay during optimization. Because interference varies across incoming domains, a constant replay coefficient may provide insufficient protection when forgetting increases or unnecessarily constrain adaptation when previous-task performance remains stable. Motivated by adaptive knowledge-preservation strategies \cite{zhu2024boosting,liu2025enhancing}, RAR-Net introduces forgetting-conditioned replay that monitors previous-task validation performance and adjusts replay strength using a smoothed forgetting signal.

Heterogeneous annotation protocols introduce an additional challenge. An anatomical structure without an annotation is not necessarily absent from the image, and treating unannotated regions as background (BG) can introduce incorrect supervision \cite{shi2021marginal,jiang2025labeled}. Similar background-confusion problems have been identified in continual semantic segmentation under partial category annotations \cite{yin2025beyond}. In our experimental setting, CAMUS and CardiacNet provide multi-structure supervision for LV, MYO, and LA, whereas EchoNet-Dynamic provides LV-specific annotations \cite{leclerc2019deep,ouyang2020video,yang2024cardiacnet}. RAR-Net therefore employs structure-aware replay to preserve MYO and LA knowledge using fully annotated historical examples. An asymmetric Tversky objective \cite{Salehi2017Tversky} emphasizes vulnerable structures while avoiding supervision of unavailable labels in partially annotated samples.

These challenges motivate a unified framework that coordinates exemplar reliability, protected historical memory, adaptive replay strength, and anatomical preservation. Unlike approaches that primarily address individual aspects of continual knowledge retention \cite{perkonigg2021dynamic,li2022domain,ranem2024atlas,zhu2024boosting,yin2025beyond}, RAR-Net integrates these mechanisms within a fixed shared encoder-decoder architecture with segmentation and auxiliary boundary heads. Its central contribution is the coordinated management of historical knowledge under sequential domain and supervision shifts, rather than architectural expansion.

The main contributions of this work are summarized as follows:

\begin{itemize}

    \item We formulate continual echocardiographic segmentation under sequential domain and supervision shifts using CAMUS, CardiacNet, and EchoNet-Dynamic, incorporating both multi-structure and LV-only annotations.

    \item We develop reliability-aware protected replay memory that combines class, boundary, image, anatomical, and temporal evidence for exemplar prioritization while maintaining historical-domain representation through task-specific memory partitions.

    \item We introduce forgetting-conditioned adaptive replay that regulates historical supervision using smoothed previous-task performance degradation, replacing fixed replay weighting with feedback-driven adaptation.

    \item We propose structure-aware replay using asymmetric Tversky penalties to preserve MYO and LA knowledge on fully annotated historical examples without treating unavailable annotations as background.

\end{itemize}


\section{Related Work}
\label{relatework}

Table~\ref{tab:related_work_comparison} summarizes representative CL approaches and their differences from RAR-Net, focusing on memory construction, knowledge preservation, and adaptation under heterogeneous domains and annotations.

\begin{table}[!t]
\centering
\caption{Comparison of CL approaches with RAR-Net
in memory construction, knowledge preservation, and domain adaptation.}
\label{tab:related_work_comparison}

\small
\setlength{\tabcolsep}{3pt}
\renewcommand{\arraystretch}{1.05}

\begin{tabularx}{\linewidth}{
    @{}
    >{\hsize=0.70\hsize\linewidth=\hsize
      \raggedright\arraybackslash}X
    >{\hsize=0.80\hsize\linewidth=\hsize
      \raggedright\arraybackslash}X
    >{\hsize=1.50\hsize\linewidth=\hsize
      \raggedright\arraybackslash}X
    @{}
}
\toprule
\textbf{Category} &
\textbf{Representative methods} &
\textbf{Comparison with RAR-Net} \\
\midrule

Regularization &
EWC \cite{kirkpatrick2017overcoming} &
Protects parameters using Fisher information;
RAR-Net preserves exemplars and anatomy. \\

Distillation &
LwF \cite{li2017learning};
\cite{zhu2024boosting} &
Distills previous-model knowledge;
RAR-Net uses reliability-guided, adaptive replay. \\

Exemplar replay &
iCaRL \cite{rebuffi2017icarl} &
Stores exemplars; RAR-Net adds reliability-based
selection and protected task memory. \\

Logit replay &
DER \cite{buzzega2020dark} &
Matches stored logits; RAR-Net emphasizes exemplar
reliability and forgetting-adaptive replay. \\

Interference replay &
MIR \cite{aljundi2019online} &
Retrieves samples vulnerable to interference;
RAR-Net uses reliability and observed forgetting. \\

Dynamic memory &
 \cite{perkonigg2021dynamic} &
Uses style and pseudo-domain diversity;
RAR-Net adds anatomical and temporal reliability. \\

Style replay &
 \cite{li2022domain} &
Uses style reconstruction and feature whitening;
RAR-Net selects exemplars across domain and annotation shifts. \\

Generative replay &
GarDA \cite{chen2023generative} &
Generates historical appearance;
RAR-Net stores bounded real exemplars. \\

Feature replay &
FOCUS \cite{Joham2025FOCUS} &
Replays features; RAR-Net stores image--annotation
pairs for spatial and anatomical supervision. \\

\textbf{Reliability-aware replay} &
\textbf{RAR-Net (ours)} &
\textbf{Combines reliability-ranked exemplars, protected
task memory, adaptive replay, and annotation-aware
structure preservation.} \\

\bottomrule
\end{tabularx}
\end{table}

\subsection{Cardiac Image Segmentation}
\label{sec:cardiac_segmentation}

Automated cardiac segmentation supports quantitative assessment of cardiovascular anatomy and function but remains challenging due to image artifacts, ambiguous boundaries, and variability across acquisition conditions. These challenges have motivated advances in convolutional neural networks, Transformer-based architectures, and medical foundation models \cite{huang2026multi,yu2025fm2,niu2025safedhdm}.

U-Net \cite{ronneberger2015u} established encoder-decoder segmentation with skip connections, enabling effective integration of semantic and spatial information.  \cite{leclerc2019deep} demonstrated multi-structure echocardiographic segmentation using CAMUS, while \cite{ouyang2020video} extended automated analysis to large-scale cardiac ultrasound video through EchoNet-Dynamic. These studies established important segmentation benchmarks but did not specifically address knowledge retention during sequential domain adaptation.

Transformer-based approaches such as TransUNet \cite{chen2024transunet} incorporate attention mechanisms to capture long-range dependencies while preserving spatial localization. Medical foundation models, including MedSAM \cite{ma2024segment}, further improve generalization through large-scale heterogeneous training. Nevertheless, representational capacity and cross-domain generalization are distinct from CL: a model may generalize to unseen data yet still forget previously acquired knowledge when updated sequentially. Consequently, advances in segmentation architectures do not eliminate the need for explicit knowledge-preservation mechanisms.

\subsection{Continual Learning}
\label{sec:continual_learning}

CL enables sequential model adaptation while mitigating catastrophic forgetting. Existing CL strategies include regularization, architectural adaptation, knowledge distillation, and replay \cite{parisi2019continual,de2021continual}. 
Elastic Weight Consolidation (EWC) protects parameters considered important for previous tasks \cite{kirkpatrick2017overcoming}, whereas Learning without Forgetting (LwF) preserves earlier model responses through distillation \cite{li2017learning}.
Recently, \cite{zhu2024boosting} introduced tri-enhanced distillation for domain-continual medical segmentation. 
Architectural approaches reduce interference through task-specific capacity but may increase model complexity. Replay-based methods, including iCaRL \cite{rebuffi2017icarl}, retain historical examples to preserve previously learned knowledge. For medical image segmentation, replay provides direct access to spatial annotations, but limited memory capacity and heterogeneous domains make exemplar selection and allocation critical. These challenges motivate reliability-aware memory management and adaptive replay strategies.

Dark Experience Replay (DER), \cite{buzzega2020dark}, extends conventional rehearsal by preserving historical model logits and using them to regularize subsequent updates, thereby reducing
interference with previously learned representations. \cite{aljundi2019online} proposed
Maximally Interfered Retrieval (MIR), which selects replay examples according to their anticipated loss increase
following a model update. These methods demonstrate the importance of preserving historical predictions and
prioritizing examples vulnerable to interference. However, neither explicitly addresses anatomical reliability, task-aware memory protection, or heterogeneous segmentation annotations, which are central considerations in RAR-Net.

\paragraph{Generative replay and gradient-constrained learning.}
~\cite{shin2017continual} introduced Deep Generative Replay, which uses a generative model to
reconstruct historical training distributions without explicitly storing previous examples. In contrast,
Gradient Episodic Memory (GEM)~\cite{lopez2017gradient} retains a small episodic memory and constrains parameter
updates to limit interference with previously learned tasks. ~\cite{elhoseiny2019efficient}
subsequently proposed Averaged GEM (A-GEM), reducing the computational cost of gradient projection through
an averaged reference gradient. These approaches illustrate different strategies for preserving historical
knowledge through generated samples or gradient-level constraints. RAR-Net instead retains annotated
echocardiographic examples in protected task-specific memory and adjusts their contribution using
forgetting-conditioned replay. Unlike GEM and A-GEM, RAR-Net does not impose explicit gradient constraints.
Its memory selection additionally considers image, boundary, class, anatomical, and temporal characteristics
to support dense segmentation across heterogeneous domains. However, retaining real images introduces
storage and data-governance requirements that generative replay can potentially reduce.

\paragraph{Continual segmentation under incomplete supervision.}
~\cite{douillard2021plop} proposed PLOP for continual semantic segmentation, combining
knowledge distillation with pseudo-labeling to mitigate forgetting and address background-label ambiguity
when previously learned categories are no longer explicitly annotated. This problem is closely related
to continual echocardiographic segmentation, where later datasets may provide annotations for only a
subset of previously learned cardiac structures. However, RAR-Net follows a different preservation
strategy. Rather than relying primarily on pseudo-labels and distillation, it replays fully
annotated historical images and applies structure-specific asymmetric Tversky losses to
reinforce myocardium and left atrium segmentation during subsequent LV-only training. This approach
uses available historical ground-truth annotations to maintain supervision for structures absent from
the current dataset. Nevertheless, it requires access to retained annotated examples and does
not explicitly guarantee anatomical topology.

\subsection{Continual Medical Segmentation}
\label{sec:continual_medical_segmentation}

Continual medical segmentation must preserve not only semantic classes but also spatial boundaries and anatomical relationships across changing domains. Domain shifts may arise from scanners, institutions, acquisition protocols, and annotation conventions, while limited historical data and heterogeneous supervision complicate knowledge retention \cite{wu2024continual,kumari2025continual}.  \cite{perkonigg2021dynamic} proposed dynamic memory that identifies pseudo-domains from image style and maintains diverse historical examples. Their experiments on multi-vendor cardiac MRI demonstrated the value of diversity-aware memory for reducing forgetting. However, appearance diversity does not necessarily indicate anatomical importance; a visually representative sample may contain limited information about vulnerable structures or difficult boundaries.
 \cite{li2022domain} introduced style-oriented replay and domain-sensitive feature whitening for domain-incremental cardiac segmentation. Their method reconstructs previous-domain appearance while reducing reliance on domain-specific features. Although effective for appearance shifts, it does not explicitly prioritize real historical examples according to anatomical reliability or address the heterogeneous supervision considered in this study. \cite{chen2023generative} proposed Generative Appearance Replay for continual Domain Adaptation (GarDA), enabling sequential adaptation without retaining historical-domain images. Evaluated on optic-disc, cardiac, and prostate segmentation, this approach reduces dependence on stored patient data. However, preservation relies on generated appearance, which may introduce reconstruction errors, and its continual unsupervised domain-adaptation setting differs from selective replay under heterogeneous supervised annotations.

\cite{zhu2024boosting} developed tri-enhanced distillation for domain-continual medical segmentation by combining stochastic knowledge augmentation, adaptive transfer, and uncertainty-guided fusion. Their findings emphasize that historical knowledge should be transferred selectively rather than uniformly. Nevertheless, the method relies on teacher-student distillation rather than protected exemplar memory or replay strength directly regulated by measured forgetting. Feature and latent replay preserve intermediate representations instead of complete images, potentially reducing memory requirements and exposure of raw patient data \cite{kumari2025continual}. However, previously stored features may become inconsistent as the encoder evolves, and compressed representations can lose fine-grained spatial information required for anatomical segmentation.

These studies demonstrate complementary approaches to historical knowledge preservation, including appearance diversity, generative reconstruction, adaptive distillation, and feature replay. However, most address individual aspects of CL and commonly assume compatible annotation spaces. In our setting, CAMUS and CardiacNet provide LV, MYO, and LA supervision, whereas EchoNet-Dynamic provides LV-only annotations. This introduces an additional requirement: preserving previously learned structures without interpreting unavailable annotations as background.

\section{Methodology}
\label{sec:methodology}

\subsection{Problem Formulation}
\label{sec:problem_formulation}

We formulate cardiac ultrasound segmentation as a sequential CL problem in which a single model encounters a sequence of datasets without unrestricted access to previously observed training data.
Let

\begin{equation}
\mathcal{T}
=
\left\{
\mathcal{D}_{1},
\mathcal{D}_{2},
\ldots,
\mathcal{D}_{T}
\right\}
\end{equation}

denote a sequence of $T$ segmentation tasks. In the primary experimental
protocol,

\begin{equation}
\mathcal{D}_{1}=\mathcal{D}_{\mathrm{CAMUS}},\qquad
\mathcal{D}_{2}=\mathcal{D}_{\mathrm{CardiacNet}},\qquad
\mathcal{D}_{3}=\mathcal{D}_{\mathrm{EchoNet}}.
\label{eq:task_sequence}
\end{equation}

Each task contains a set of image-annotation pairs,

\begin{equation}
\mathcal{D}_{t}
=
\left\{
(\mathbf{x}_{i}^{t},
 \mathbf{y}_{i}^{t},
 m_i^t)
\right\}_{i=1}^{N_t},
\end{equation}

where $\mathbf{x}_{i}^{t}\in\mathbb{R}^{H\times W}$ denotes an echocardiographic image, $\mathbf{y}_{i}^{t}$ is its segmentation mask, and $m_i^t$ denotes the annotation availability associated with the sample. The latter is necessary because the supervision is heterogeneous across the sequence. CAMUS and CardiacNet contain annotations for the LV, MYO, and LA, whereas EchoNet-Dynamic provides LV supervision only. The segmentation model is represented by

\begin{equation}
f_{\theta}:
\mathbb{R}^{H\times W}
\rightarrow
\mathbb{R}^{C\times H\times W},
\end{equation}

where $\theta$ denotes the learnable parameters and $C=4$
corresponds to BG, LV, MYO, and LA. For an input image $\mathbf{x}$, the model produces logits

\begin{equation}
\mathbf{z}=f_{\theta}(\mathbf{x}),
\end{equation}

which are converted to class probabilities using

\begin{equation}
p_{c}(u)
=
\frac{\exp(z_c(u))}
{\sum_{k=1}^{C}\exp(z_k(u))},
\qquad c=1,\ldots,C,
\label{eq:softmax}
\end{equation}

for pixel $u$. The predicted segmentation is

\begin{equation}
\hat{\mathbf{y}}(u)
=
\arg\max_{c}p_c(u).
\end{equation}

At task $t$, the learner has full access only to $\mathcal{D}_t$. Historical
information is retained through a bounded replay memory

\begin{equation}
\mathcal{M}_{t}
=
\bigcup_{k=1}^{t}
\mathcal{M}_{t}^{(k)},
\qquad
|\mathcal{M}_{t}|\leq M,
\label{eq:memory_constraint}
\end{equation}

where $M=1800$ in our experiments and
$\mathcal{M}_{t}^{(k)}$ denotes the protected memory partition associated
with task $k$. The objective is therefore not merely to minimize the risk on the currently
observed distribution,

\begin{equation}
\min_{\theta}
\mathbb{E}_{(\mathbf{x},\mathbf{y})\sim\mathcal{D}_{t}}
\left[
\mathcal{L}
\left(
f_{\theta}(\mathbf{x}),\mathbf{y}
\right)
\right],
\end{equation}

because such optimization can overwrite parameters required by earlier
domains. Instead, we seek parameters $\theta_t$ that minimize performance
degradation across the entire observed task history,

\begin{equation}
\theta_t^{*}
=
\arg\min_{\theta}
\sum_{k=1}^{t}
\omega_k
\mathcal{R}_{k}(\theta),
\label{eq:cl_objective}
\end{equation}

where $\mathcal{R}_{k}$ represents the segmentation risk associated with task $k$ and $\omega_k$ controls its effective contribution. Since historical datasets are no longer fully available, RAR-Net approximates their
contribution using reliability-aware protected replay.

\subsection{Overall RAR-Net Framework}
\label{sec:overall_framework}

\begin{figure*}[!t] 
\centering 
\includegraphics[width=\textwidth]{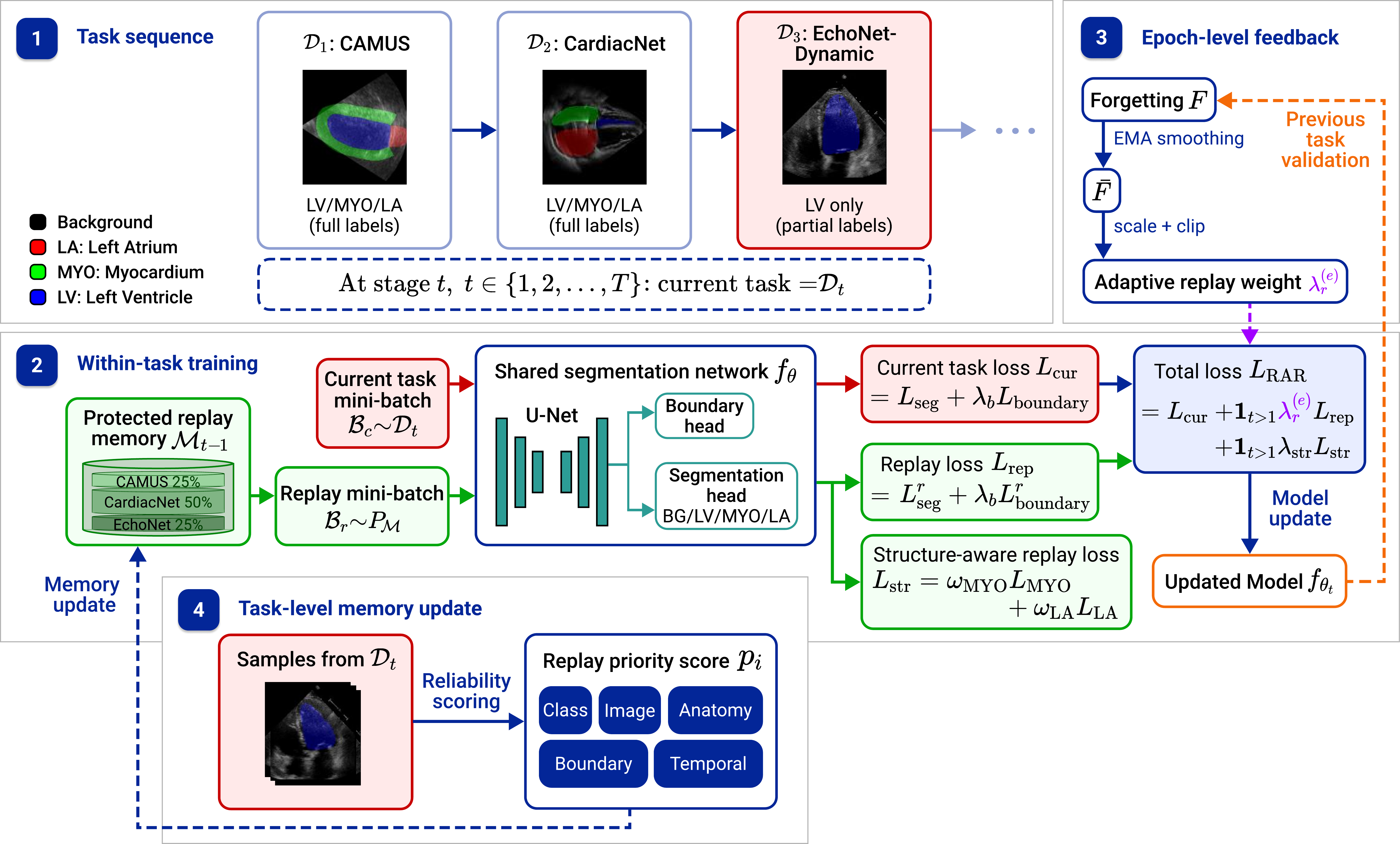} 
\caption{Overview of the RAR-Net framework for continual echocardiographic segmentation. The model sequentially learns CAMUS, CardiacNet, and EchoNet-Dynamic using a shared segmentation network and protected replay memory. Training combines current-task, replay, and structure-aware losses, while an exponential moving average (EMA) of previous-task forgetting adaptively controls the replay weight $\lambda_r^{(e)}$. Historical samples are prioritized using class, boundary, image, anatomical, and temporal reliability factors. Memory allocation follows predefined task weights normalized over the observed tasks.}
\label{Framwork} 
\end{figure*}

The overall architecture and CL workflow of RAR-Net are illustrated in Figure~\ref{Framwork}. As shown, RAR-Net addresses sequential echocardiographic segmentation through four interacting mechanisms: (i) a shared encoder-decoder segmentation network with segmentation and boundary prediction heads, (ii) reliability-aware protected task memory for preserving informative historical examples, (iii) forgetting-conditioned adaptive replay that dynamically regulates the contribution of historical knowledge, and (iv) structure-aware replay for preserving cardiac anatomy under heterogeneous annotation. These components operate jointly to balance adaptation to incoming domains with retention of previously acquired segmentation knowledge. The primary learning trajectory is, CAMUS $\longrightarrow CardiacNet \longrightarrow EchoNet\-Dynamic.$


After learning task $\mathcal{D}_{t}$, selected examples are inserted into $\mathcal{M}$ according to their reliability scores and the protected quota assigned to their source task. When task $t>1$ is learned, a replay mini-batch

\begin{equation}
\mathcal{B}_{r}
\sim
P_{\mathcal{M}}
\end{equation}

is sampled from memory together with the current-task mini-batch

\begin{equation}
\mathcal{B}_{c}
\sim
\mathcal{D}_{t}.
\end{equation}

Historical samples are therefore not treated as an undifferentiated archive. RAR-Net first determines which examples deserve storage, then protects their task representation, and finally adjusts their influence on optimization according to measured forgetting. Conceptually, the complete framework can be expressed as

\begin{equation}
\text{Current Task}
+\underbrace{\text{Reliable Historical Evidence}}_{\text{protected memory}}
+\underbrace{\text{Forgetting Feedback}}_{\text{adaptive replay}}
+\underbrace{\text{Anatomical Constraints}}_{\text{structure replay}}
\longrightarrow \theta_t.
\end{equation}

\subsection{Segmentation Network Architecture}
\label{sec:architecture}

RAR-Net uses a U-Net-style encoder-decoder as its segmentation backbone. The architectural design is deliberately kept compact so that improvements in continual performance can be attributed to the replay mechanisms rather than to progressive network expansion. Given an input image

\begin{equation}
\mathbf{x}\in\mathbb{R}^{1\times256\times256},
\end{equation}

the encoder progressively transforms the image into hierarchical feature
representations,

\begin{equation}
\mathbf{h}_{l}
=
E_l(\mathbf{h}_{l-1}),
\qquad
l=1,\ldots,L,
\end{equation}

where $\mathbf{h}_0=\mathbf{x}$ and $E_l(\cdot)$ represents the $l$-th encoder block. The base channel width is 32 and increases with network depth. The decoder reconstructs spatial resolution through successive upsampling operations. At decoder level $l$, the corresponding encoder representation is transferred through a skip connection,

\begin{equation}
\mathbf{d}_{l}
=
D_l
\left(
\left[
U(\mathbf{d}_{l+1});
\mathbf{h}_{l}
\right]
\right),
\label{eq:skip_connection}
\end{equation}

where $U(\cdot)$ denotes upsampling, $[\cdot;\cdot]$ denotes channel-wise concatenation, and $D_l(\cdot)$ denotes the decoder transformation. The final decoder representation $\mathbf{d}_1$ is mapped to segmentation logits through the segmentation head,

\begin{equation}
\mathbf{z}_{\mathrm{seg}}
=
g_{\mathrm{seg}}(\mathbf{d}_1).
\end{equation}

A boundary prediction branch additionally maps the shared decoder representation to a boundary estimate,

\begin{equation}
\mathbf{z}_{b}
=
g_{b}(\mathbf{d}_1).
\end{equation}

The two heads therefore share the same anatomical representation while optimizing complementary region and contour information. The architecture contains approximately $7.76$ million trainable parameters and remains fixed throughout the complete CL sequence.

\subsection{Reliability-Aware Protected Task Memory}
\label{sec:protected_memory}

\begin{figure}[!t] 
\centering 
\includegraphics[width=1\textwidth]{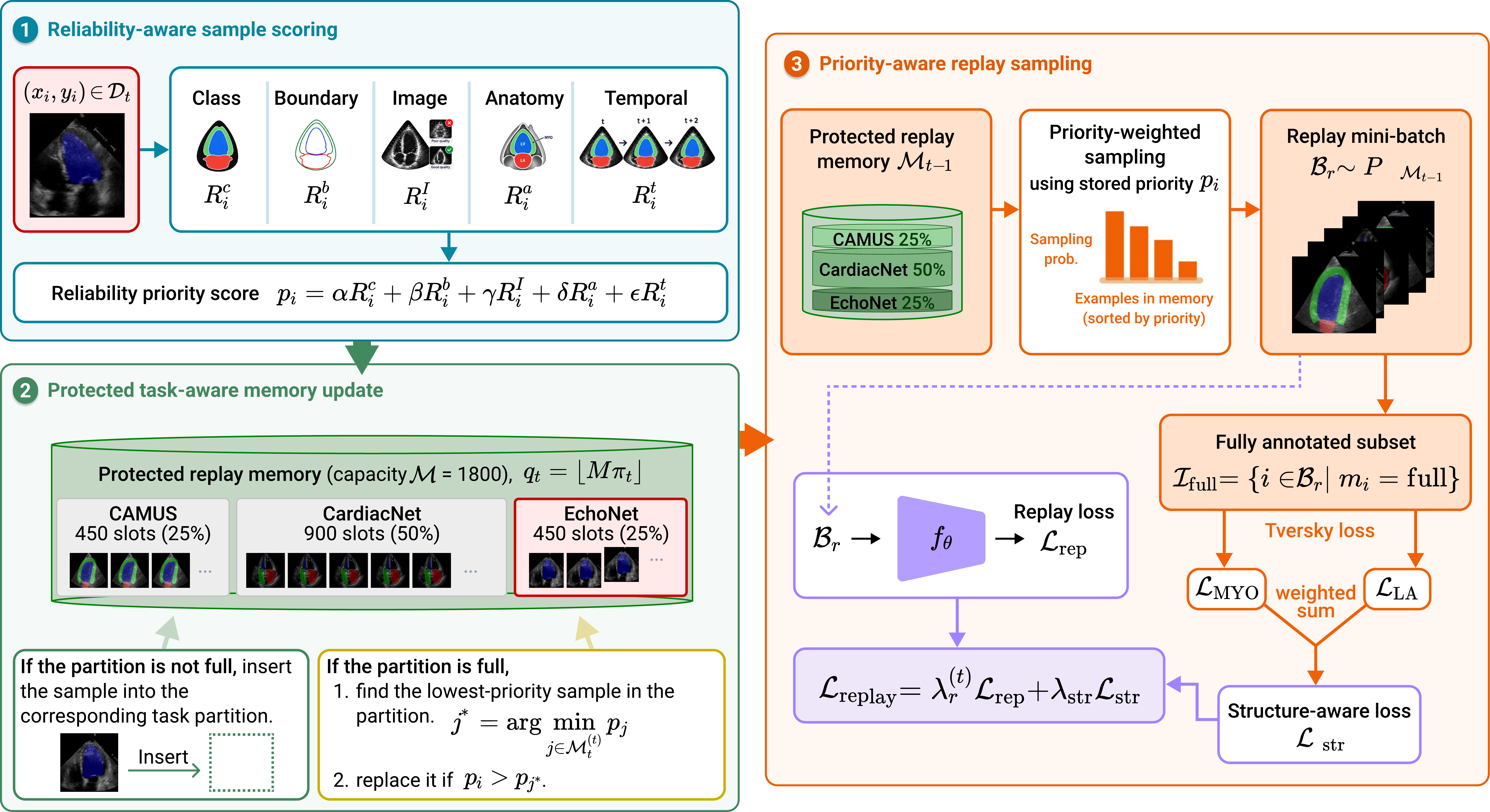} 
\caption{Reliability-aware protected memory mechanism of RAR-Net. Candidate samples are evaluated using complementary class, boundary, image, anatomical, and temporal reliability factors, which are combined into a unified priority score. Samples are subsequently maintained within task-specific protected memory partitions to limit domination by newly observed domains. The illustrated 450/900/450 partition represents the protected-memory configuration after all three tasks have been observed; at earlier stages, the predefined relative task-allocation weights are renormalized over the tasks observed up to that point to determine the corresponding memory quotas. Priority-aware sampling retrieves representative previous-task examples for subsequent replay learning.}
\label{rarnet} 
\end{figure}

Figure~\ref{rarnet} illustrates the reliability-aware sample scoring, protected task-aware memory update, and priority-aware replay sampling mechanism. A fixed replay budget creates a selection problem: not every previously observed sample can be retained, and preserving an  uninformative example necessarily reduces the capacity available for potentially more useful historical information. RAR-Net therefore estimates the replay value of each candidate sample before memory insertion using complementary information from its annotation, image appearance, anatomical geometry, and temporal context. For sample $i$, the reliability representation is defined as

\begin{equation}
\mathbf{r}_i
=
\left[
R_i^{c},
R_i^{b},
R_i^{I},
R_i^{a},
R_i^{t}
\right],
\end{equation}

where $R_i^{c}$, $R_i^{b}$, $R_i^{I}$, $R_i^{a}$, and $R_i^{t}$ denote class, boundary, image, anatomical, and temporal reliability, respectively. All five components are constrained to the interval $[0,1]$ before fusion. Importantly, these quantities are computed directly from the normalized input image, available ground-truth (GT) annotation, and dataset-specific temporal information. Model prediction uncertainty is not used in the final RAR-Net configuration.

\paragraph{Class reliability.}
Class reliability measures the amount of foreground-class supervision available in an annotated sample. Let $\mathcal{C}_{fg}=\{1,\ldots,C-1\}$
denote the foreground classes and let $\mathbbm{1}[\cdot]$ be the indicator function. It is computed as

\begin{equation}
R_i^{c}
=
\frac{1}{C-1}
\sum_{c=1}^{C-1}
\mathbbm{1}
\left[
\exists\,u:\, y_i(u)=c
\right],
\label{eq:class_reliability}
\end{equation}

where $y_i$ denotes the GT segmentation mask and $u$ indexes image pixels. Thus, samples containing a larger fraction of the available foreground classes receive a higher class-content score.

\paragraph{Boundary reliability.}
Boundary reliability quantifies the amount of structural boundary information contained in the GT mask. A binary boundary map $\mathcal{B}(y_i)$ is generated by identifying label transitions within a $3\times3$ neighborhood. The boundary score is then defined as the fraction of boundary pixels,

\begin{equation}
R_i^{b}
=
\frac{1}{HW}
\sum_{u}
\mathcal{B}(y_i)(u),
\label{eq:boundary_reliability}
\end{equation}

where $H$ and $W$ denote the image height and width. Since $\mathcal{B}(y_i)\in\{0,1\}^{H\times W}$, this score is naturally bounded within $[0,1]$. A larger value indicates that the sample contains more boundary information that may be useful for preserving spatial delineation during replay. 

\paragraph{Image reliability.}
Image reliability is computed from the normalized input image $x_i\in[0,1]$ using three complementary image-quality cues: intensity contrast, local sharpness, and foreground visibility. Specifically,

\begin{align}
R_i^{\mathrm{con}}
&=
\operatorname{clip}
\left(
\frac{\sigma(x_i)}{0.25},0,1
\right),\\
R_i^{\mathrm{sharp}}
&=
\operatorname{clip}
\left(
\frac{\operatorname{Var}(\nabla^2 x_i)}{500},0,1
\right),\\
R_i^{\mathrm{vis}}
&=
\operatorname{clip}
\left(
\frac{
\frac{1}{HW}\sum_u
\mathbbm{1}[x_i(u)>0.05]
}{0.60},
0,1
\right),
\end{align}

where $\sigma(\cdot)$ denotes image intensity standard deviation and $\nabla^2$ denotes the Laplacian operator. The image reliability is

\begin{equation}
R_i^{I}
=
0.4R_i^{\mathrm{con}}
+
0.4R_i^{\mathrm{sharp}}
+
0.2R_i^{\mathrm{vis}}.
\label{eq:image_reliability}
\end{equation}

The final value is clipped to $[0,1]$. This formulation favors samples with sufficient contrast, locally defined image structure, and adequate visible image content.

\paragraph{Anatomical reliability.}
Anatomical reliability is estimated from the union of foreground structures in the GT mask. Let $A_i$ denote foreground area, $P_i$ the boundary-pixel count, and

\begin{equation}
r_i^{A}=\frac{A_i}{HW}.
\end{equation}

Samples containing fewer than 10 foreground pixels are assigned $R_i^{a}=0$. Otherwise, an area-plausibility term is defined as

\begin{equation}
R_i^{\mathrm{area}}
=
\begin{cases}
1, & 0.005 \le r_i^{A}\le0.55,\\
0.4, & \text{otherwise},
\end{cases}
\end{equation}

while compactness and boundary smoothness are calculated as

\begin{align}
R_i^{\mathrm{comp}}
&=
\operatorname{clip}
\left(
\frac{4\pi A_i}{P_i^2+\varepsilon},
0,1
\right),\\
R_i^{\mathrm{smooth}}
&=
\operatorname{clip}
\left(
\frac{1}
{1+P_i/(\sqrt{A_i}+\varepsilon)},
0,1
\right).
\end{align}

The anatomical reliability is consequently

\begin{equation}
R_i^{a}
=
\operatorname{clip}
\left(
0.4R_i^{\mathrm{area}}
+
0.4R_i^{\mathrm{comp}}
+
0.2R_i^{\mathrm{smooth}},
0,1
\right).
\label{eq:anatomical_reliability}
\end{equation}

This term favors masks with sufficient foreground support and comparatively plausible geometric characteristics.

\paragraph{Temporal reliability.}
Temporal reliability depends on whether temporal information is available for the corresponding dataset. For EchoNet-Dynamic, temporal reliability is derived from the variation in LV area across the available traced frames of each video. Let $A_{\max}$ and $A_{\min}$ denote the maximum and minimum LV areas among the traced frames. For videos containing valid LV regions,

\begin{equation}
r_i^{t}
=
\frac{A_{\min}}{A_{\max}+\varepsilon}.
\end{equation}

The temporal score is assigned as

\begin{equation}
R_i^{t}
=
\begin{cases}
0.1, & A_i\le10,\\
1.0, & 0.20\le r_i^{t}\le0.95,\\
0.5, & \text{otherwise}.
\end{cases}
\label{eq:temporal_reliability}
\end{equation}

When fewer than two traced frames are available, a neutral value of $0.5$ is used. CAMUS and CardiacNet are processed as independent annotated slices in the present implementation and therefore do not provide an equivalent frame-wise temporal estimate; a neutral temporal reliability of $0.5$ is assigned to their samples. Hence, the temporal term contributes dataset-specific dynamic information only where such information is available, rather than implying identical temporal modeling across all three datasets. The overall replay priority is then obtained by reliability fusion,

\begin{equation}
p_i
=
\alpha R_i^{c}
+
\beta R_i^{b}
+
\gamma R_i^{I}
+
\delta R_i^{a}
+
\epsilon R_i^{t},
\label{eq:priority}
\end{equation}

subject to

\begin{equation}
\alpha+\beta+\gamma+\delta+\epsilon=1.
\end{equation}

In the final RAR-Net configuration,

\begin{equation}
(\alpha,\beta,\gamma,\delta,\epsilon)
=
(0.15,0.20,0.20,0.25,0.20).
\end{equation}

Because every component lies in $[0,1]$ and the fusion coefficients are non-negative and sum to one, $p_i\in[0,1]$. The final configuration assigns the largest individual coefficient to anatomical reliability, thereby giving slightly greater influence to structural plausibility during reliability fusion.

A global reliability ranking alone can nevertheless allow a large or recent task to dominate the replay memory. RAR-Net therefore combines reliability ranking with protected task-aware memory. Rather than assigning fixed capacities to all tasks from the beginning, the predefined task fractions are treated as relative allocation weights and dynamically normalized over the tasks observed up to the current learning stage.

Let $\mathcal{T}_{\mathrm{seen}}$ denote the set of tasks encountered so far.
For an observed task $t\in\mathcal{T}_{\mathrm{seen}}$, its protected quota is

\begin{equation}
q_t
=
\operatorname{round}
\left(
M
\frac{\pi_t}
{\sum_{k\in\mathcal{T}_{\mathrm{seen}}}\pi_k}
\right),
\label{eq:quota}
\end{equation}

where $M$ is the total replay-memory capacity and $\pi_t$ is the predefined
relative allocation weight for task $t$. Because integer rounding can cause
the resulting quotas to differ slightly from $M$, the quotas are adjusted
when necessary such that

\begin{equation}
\sum_{t\in\mathcal{T}_{\mathrm{seen}}} q_t = M.
\end{equation}

For the three-domain sequence, the final RAR-Net configuration uses

\begin{equation}
\pi_{\mathrm{CAMUS}}=0.25,\qquad
\pi_{\mathrm{CardiacNet}}=0.50,\qquad
\pi_{\mathrm{EchoNet}}=0.25.
\end{equation}

These values represent relative task-protection weights rather than fixed
capacities throughout the entire learning sequence. With a total memory
capacity of $M=1800$, when CAMUS is the only observed task, its normalized
quota is

\begin{equation}
q_{\mathrm{CAMUS}}=1800.
\end{equation}

After CardiacNet is introduced, the fractions are renormalized over CAMUS
and CardiacNet, yielding

\begin{equation}
q_{\mathrm{CAMUS}}=600,\qquad
q_{\mathrm{CardiacNet}}=1200.
\end{equation}

After EchoNet-Dynamic is introduced, all three predefined fractions
participate in the normalization and the protected quotas become

\begin{equation}
q_{\mathrm{CAMUS}}=450,\qquad
q_{\mathrm{CardiacNet}}=900,\qquad
q_{\mathrm{EchoNet}}=450.
\end{equation}

Thus, the protected allocation evolves as new tasks enter the CL stream. This dynamic normalization allows the available replay
capacity to be fully utilized during early stages while converging to the
intended $25\%, 50\%, 25\%$ allocation once all three domains have been
observed. During memory construction, candidate samples are initially inserted while
the replay buffer has available capacity. Once the memory reaches capacity,
the protected quotas govern replacement. For a candidate $i$ belonging to
task $t$, if task $t$ is underrepresented relative to its current quota,
replacement is sought from an overrepresented task. Otherwise, the candidate
competes with stored samples belonging to the same task. In either case, the
lowest-priority eligible exemplar is replaced only when the incoming sample
has a higher reliability priority.

Consequently, the two mechanisms play complementary roles: the normalized
task quotas determine how the finite replay capacity is distributed across
the observed domains, whereas the reliability priority determines which
examples are preferentially retained within those allocations.

\subsection{Forgetting-Conditioned Adaptive Replay}
\label{sec:adaptive_replay}

Fixed replay strength assumes that historical knowledge requires the same degree of protection throughout training. This assumption is restrictive: forgetting can remain weak during one stage of optimization and increase sharply after a substantial domain shift. RAR-Net therefore adapts replay intensity using validation-based forgetting feedback evaluated after each
training epoch. Let $S_k^{\mathrm{best}}$ denote the best validation Dice previously observed for an earlier task $k$, and let $S_k^{(e)}$ denote its validation Dice at the end of epoch $e$ while learning the current task $t$. For $t>1$, the instantaneous forgetting signal is computed over the previously learned tasks as

\begin{equation}
F^{(e)}
=
\frac{1}{t-1}
\sum_{k<t}
\max
\left(
0,
S_k^{\mathrm{best}}-S_k^{(e)}
\right).
\label{eq:forgetting}
\end{equation}

Only performance degradation contributes to the controller; improvements on earlier tasks do not artificially reduce the estimated need for replay. To reduce sensitivity to transient validation fluctuations, RAR-Net maintains an exponential moving average (EMA) of the forgetting signal,

\begin{equation}
\bar{F}^{(e)}
=
\rho \bar{F}^{(e-1)}
+
(1-\rho)F^{(e)},
\label{eq:ema}
\end{equation}

where $\rho=0.70$.

The controller is updated at the end of each epoch. After the warm-up period of $E_w=3$ epochs, the replay coefficient used in the subsequent epoch is

\begin{equation}
\lambda_r^{(e+1)}
=
\operatorname{clip}
\left(
\lambda_r+\kappa\bar{F}^{(e)},
\lambda_r^{\min},
\lambda_r^{\max}
\right),
\label{eq:adaptive_lambda}
\end{equation}

where

\begin{equation}
\lambda_r=1.0,\qquad
\kappa=4.0,\qquad
\lambda_r^{\min}=0.75,\qquad
\lambda_r^{\max}=1.75.
\end{equation}

During the warm-up period,

\begin{equation}
\lambda_r^{(e+1)}=\lambda_r,
\qquad e\leq E_w.
\end{equation}

Thus, validation performance on previously learned tasks is evaluated after each epoch, and the resulting forgetting estimate determines the replay strength used during the next epoch. {Equation}~\eqref{eq:adaptive_lambda} therefore implements a feedback controller in which historical information receives greater optimization weight when previous-task performance deteriorates, while the clipping interval prevents replay from overwhelming learning on the current task.

\subsection{Structure-Aware Replay}
\label{sec:structure_replay}

The most consequential difficulty in the present CL setting is not merely domain shift but annotation shift. CAMUS and CardiacNet provide LV, MYO, and LA annotations, whereas EchoNet-Dynamic provides LV supervision only. Consequently, during the final task, LV continues to receive direct gradients but MYO and LA do not. A generic replay objective can reduce this imbalance, but it does not explicitly guarantee preservation of the structures that disappear from current-task supervision. RAR-Net therefore introduces structure-aware replay for fully annotated historical samples. Let

\begin{equation}
\mathcal{I}_{\mathrm{full}}
=
\left\{
i\in\mathcal{B}_{r}
\;|\;
m_i=\mathrm{full}
\right\}
\end{equation}

denote replay samples containing complete LV/MYO/LA annotations. For structure $s\in\{\mathrm{MYO},\mathrm{LA}\}$, let $p_{i,s}(u)$ denote the predicted probability and $y_{i,s}(u)$ the binary reference mask. We define soft true-positive, false-positive, and false-negative quantities as

\begin{align}
TP_s &=
\sum_{i\in\mathcal{I}_{\mathrm{full}}}
\sum_{u}
p_{i,s}(u)y_{i,s}(u),
\\
FP_s &=
\sum_{i\in\mathcal{I}_{\mathrm{full}}}
\sum_{u}
p_{i,s}(u)\left[1-y_{i,s}(u)\right],
\\
FN_s &=
\sum_{i\in\mathcal{I}_{\mathrm{full}}}
\sum_{u}
\left[1-p_{i,s}(u)\right]y_{i,s}(u).
\end{align}

The Tversky index is then

\begin{equation}
TI_s
=
\frac{TP_s+\varepsilon}
{TP_s+\alpha_T FP_s+\beta_T FN_s+\varepsilon},
\label{eq:tversky}
\end{equation}

and the corresponding loss is

\begin{equation}
\mathcal{L}_{s}
=
1-TI_s.
\label{eq:tversky_loss}
\end{equation}

RAR-Net uses

\begin{equation}
\alpha_T=0.30,
\qquad
\beta_T=0.70.
\end{equation}

Because $\beta_T>\alpha_T$, false negatives receive greater penalty than false positives. This asymmetry is intentional. Under sequential learning, complete disappearance or severe erosion of an old anatomical structure is particularly undesirable. The structure-aware objective combines MYO and LA preservation:

\begin{equation}
\mathcal{L}_{\mathrm{str}}
=
\omega_{\mathrm{MYO}}
\mathcal{L}_{\mathrm{MYO}}
+
\omega_{\mathrm{LA}}
\mathcal{L}_{\mathrm{LA}},
\label{eq:structure_loss}
\end{equation}

with

\begin{equation}
\omega_{\mathrm{MYO}}=1.25,
\qquad
\omega_{\mathrm{LA}}=1.00.
\end{equation}

The larger MYO weight reflects its more difficult geometry. The MYO is a relatively thin structure surrounding the ventricular cavity; small contour displacements can therefore produce disproportionately large changes in its overlap. More importantly, once MYO supervision disappears from the current task, ordinary gradient updates provide no direct incentive to preserve this representation. Structure-aware replay restores such an incentive using fully annotated historical examples. The structure contribution is weighted by

\begin{equation}
\lambda_{\mathrm{str}}=0.30,
\end{equation}

yielding

\begin{equation}
\mathcal{L}_{\mathrm{SAR}}
=
\lambda_{\mathrm{str}}
\mathcal{L}_{\mathrm{str}}.
\end{equation}

This term should be distinguished from generic region replay. Its purpose is not to relearn all historical labels equally, but to explicitly preserve anatomical information that becomes weakly or entirely unsupervised later in the sequence.

\subsubsection{Segmentation and Boundary Supervision}
\label{sec:segmentation_loss}

RAR-Net accommodates two annotation modes: full multi-structure supervision for CAMUS and CardiacNet, and LV-only supervision for EchoNet-Dynamic. Accordingly, the segmentation objective is conditioned on the annotation mode of each sample rather than treating unavailable structures as background. For a fully annotated sample, let $\mathbf{z}_i(u)\in\mathbb{R}^{C}$ denote the segmentation logits at pixel $u$, and let $y_i(u)\in\{0,\ldots,C-1\}$ denote the corresponding GT class. The segmentation loss combines multiclass cross-entropy and foreground multiclass Dice loss:

\begin{equation}
\mathcal{L}_{\mathrm{seg}}^{\mathrm{full}}
=
\mathcal{L}_{\mathrm{CE}}
+
\mathcal{L}_{\mathrm{Dice}}^{\mathrm{multi}}.
\label{eq:full_seg_loss}
\end{equation}

The cross-entropy term is

\begin{equation}
\mathcal{L}_{\mathrm{CE}}
=
-\frac{1}{N}
\sum_{i,u}
\log
p_{i,y_i(u)}(u),
\end{equation}

where

\begin{equation}
p_{i,c}(u)
=
\frac{\exp(z_{i,c}(u))}
{\sum_{k=0}^{C-1}\exp(z_{i,k}(u))}
\end{equation}

is the softmax probability for class $c$.

The multiclass Dice loss is evaluated over the foreground classes only:

\begin{equation}
\mathcal{L}_{\mathrm{Dice}}^{\mathrm{multi}}
=
1-
\frac{1}{C-1}
\sum_{c=1}^{C-1}
\frac{
2\sum_{i,u}p_{i,c}(u)y_{i,c}(u)+\varepsilon
}{
\sum_{i,u}p_{i,c}(u)
+
\sum_{i,u}y_{i,c}(u)
+
\varepsilon
},
\label{eq:multiclass_dice}
\end{equation}

where $y_{i,c}$ is the one-hot reference mask. Background is excluded from
the Dice average.

\paragraph{LV-only partial supervision.}
EchoNet-Dynamic provides an LV annotation but does not provide MYO or LA annotations. Consequently, applying the full multiclass objective would incorrectly interpret unavailable anatomical labels as negative supervision. RAR-Net therefore uses a separate LV-only objective for these samples. Let $y_i^{\mathrm{LV}}(u)\in\{0,1\}$ denote the binary LV reference mask.
The binary LV logit used by the implementation is

\begin{equation}
\tilde{z}_i^{\mathrm{LV}}(u)
=
z_{i,\mathrm{LV}}(u)
-
z_{i,\mathrm{BG}}(u).
\label{eq:lv_binary_logit}
\end{equation}

The LV-only segmentation loss is

\begin{equation}
\mathcal{L}_{\mathrm{seg}}^{\mathrm{LV}}
=
\mathcal{L}_{\mathrm{BCE}}^{\mathrm{LV}}
+
\mathcal{L}_{\mathrm{Dice}}^{\mathrm{LV}},
\label{eq:lv_partial_loss}
\end{equation}

where

\begin{equation}
\begin{aligned}
\mathcal{L}_{\mathrm{BCE}}^{\mathrm{LV}}
= -\frac{1}{N}\sum_{i,u}
\Big[
& y_i^{\mathrm{LV}}(u)\log \sigma(\tilde{z}_i^{\mathrm{LV}}(u))
&+
(1-y_i^{\mathrm{LV}}(u))
\log(1-\sigma(\tilde{z}_i^{\mathrm{LV}}(u)))
\Big].
\end{aligned}
\label{eq:lv_bce}
\end{equation}

and

\begin{equation}
\mathcal{L}_{\mathrm{Dice}}^{\mathrm{LV}}
=
1-
\frac{
2\sum_{i,u}
p_{i,\mathrm{LV}}(u)y_i^{\mathrm{LV}}(u)
+\varepsilon
}{
\sum_{i,u}p_{i,\mathrm{LV}}(u)
+
\sum_{i,u}y_i^{\mathrm{LV}}(u)
+\varepsilon
}.
\label{eq:lv_dice}
\end{equation}

Here, $p_{i,\mathrm{LV}}$ is the LV probability obtained from the four-class softmax output. Importantly, the LV-only samples do not enter the full multiclass cross-entropy or multiclass Dice objective. Therefore, MYO and LA are not explicitly assigned background targets and receive no direct class-specific supervision from EchoNet-Dynamic. Their preservation during the LV-only stage is instead supported by replay of fully annotated historical examples and the structure-aware replay objective described in Section~\ref{sec:structure_replay}.

\paragraph{Boundary supervision.}
RAR-Net additionally predicts an auxiliary binary boundary map. For a GT segmentation $y_i$, the reference boundary $b_i(u)\in\{0,1\}$ is obtained from horizontal and vertical label transitions between neighboring pixels. The boundary objective is binary cross-entropy with logits:

\begin{equation}
\begin{aligned}
\mathcal{L}_{\mathrm{boundary}}
= -\frac{1}{N}\sum_{i,u}
\Big[
& b_i(u)\log\sigma(\hat{b}_i(u))
&+ (1-b_i(u))\log(1-\sigma(\hat{b}_i(u)))
\Big].
\end{aligned}
\label{eq:boundary_loss}
\end{equation}

where $\hat{b}_i$ denotes the boundary-head logit.

For fully annotated CAMUS and CardiacNet samples, the boundary target contains transitions among the available anatomical labels. For EchoNet-Dynamic, the reference mask is binary; hence, the boundary target corresponds only to the annotated LV contour. No MYO or LA boundary supervision is generated from EchoNet-Dynamic.

\subsection{Overall Optimization Objective}
\label{sec:optimization}

RAR-Net integrates annotation-aware segmentation, boundary supervision, adaptive replay, and structure-aware preservation within a unified training objective. For each sample $i$, the segmentation loss is selected according to its annotation mode:

\begin{equation}
\mathcal{L}_{\mathrm{seg}}^{(i)}
=
\begin{cases}
\mathcal{L}_{\mathrm{seg}}^{\mathrm{full}},
& m_i=\mathrm{full},\\
\mathcal{L}_{\mathrm{seg}}^{\mathrm{LV}},
& m_i=\mathrm{LV\mbox{-}only},
\end{cases}
\label{eq:annotation_conditioned_seg}
\end{equation}

where the full-label and LV-only losses are defined in Section~\ref{sec:segmentation_loss}. This formulation avoids assigning explicit background targets to unavailable MYO and LA annotations. The current-task and historical replay objectives are

\begin{equation}
\begin{aligned}
\mathcal{L}_{\mathrm{cur}}
&=
\mathcal{L}_{\mathrm{seg}}^{c}
+
\lambda_b\mathcal{L}_{\mathrm{boundary}}^{c},\\
\mathcal{L}_{\mathrm{rep}}
&=
\mathcal{L}_{\mathrm{seg}}^{r}
+
\lambda_b\mathcal{L}_{\mathrm{boundary}}^{r},
\end{aligned}
\label{eq:current_replay_loss}
\end{equation}

where $\lambda_b=0.30$, and the segmentation and boundary losses follow the available annotation mode for each sample. The overall optimization objective at task $t$ is

\begin{equation}
\mathcal{L}_{\mathrm{RAR}}
=
\mathcal{L}_{\mathrm{cur}}
+
\mathbbm{1}_{t>1}
\lambda_r^{(e)}\mathcal{L}_{\mathrm{rep}}
+
\mathbbm{1}_{t>1}
\lambda_{\mathrm{str}}\mathcal{L}_{\mathrm{str}},
\label{eq:total_loss}
\end{equation}

where $\mathbbm{1}_{t>1}$ activates historical replay after the first task, $\lambda_r^{(e)}$ denotes the forgetting-conditioned replay coefficient, and $\lambda_{\mathrm{str}}=0.30$ weights the structure-aware objective defined in Section~\ref{sec:structure_replay}. The latter is evaluated only on fully annotated replay samples to reinforce MYO and LA segmentation. Model parameters are optimized using Adam with a learning rate of $10^{-4}$ and weight decay of $10^{-5}$.

\subsection{Training Procedure}
\label{sec:training_algorithm}

\begin{algorithm}[!t]
\caption{Training Procedure of RAR-Net}
\label{alg:rarnet_training}
\footnotesize
\begin{algorithmic}[1]
\Require Sequential tasks $\{\mathcal{D}_1,\ldots,\mathcal{D}_T\}$,
model $f_{\theta}$, replay memory $\mathcal{M}$,
epochs $E$, replay batch size $B_r$
\Ensure Trained model $f_{\theta}$
\State Initialize $\theta$, $\mathcal{M}\leftarrow\emptyset$
\State Initialize $\bar{F}^{(0)}\leftarrow0$
\For{$t=1$ to $T$}
    \State Set current task $\mathcal{D}_t$
    \State Initialize $\lambda_r^{(1)}\leftarrow\lambda_r$
    \For{$e=1$ to $E$}
        \For{mini-batch $(X,Y,m)\sim\mathcal{D}_t$}
            \State $(\hat{Y},\hat{B})\leftarrow f_{\theta}(X)$
            \State Compute current-task loss:
            \State $\mathcal{L}_{\mathrm{cur}}=
            \mathcal{L}_{\mathrm{seg}}(X,Y,m)
            +\lambda_b\mathcal{L}_{\mathrm{bnd}}$
            \State $\mathcal{L}\leftarrow\mathcal{L}_{\mathrm{cur}}$
            \If{$\mathcal{M}\neq\emptyset$}
                \State Sample $(X_r,Y_r,m_r)\sim\mathcal{M}$
                using priority weights
                \State $(\hat{Y}_r,\hat{B}_r)\leftarrow f_{\theta}(X_r)$
                \State Compute replay loss $\mathcal{L}_{\mathrm{rep}}$
                \State $\mathcal{L}\leftarrow\mathcal{L}
                +\lambda_r^{(e)}\mathcal{L}_{\mathrm{rep}}$
                \State Select fully annotated replay samples:
                \State $\mathcal{I}_f=
                \{j:m_{r,j}=\texttt{full}\}$
                \If{$\mathcal{I}_f\neq\emptyset$}
                    \State Compute MYO and LA Tversky losses:
                    \State ${\mathcal{L}_{\mathrm{MYO}}}=
                    1-\frac{TP_m+\varepsilon}
                    {TP_m+\alpha_T FP_m+\beta_T FN_m+\varepsilon}$
                    \State ${\mathcal{L}_{\mathrm{LA}}}=
                    1-\frac{TP_a+\varepsilon}
                    {TP_a+\alpha_T FP_a+\beta_T FN_a+\varepsilon}$
                    \State Compute structure-aware replay loss:
                    \State $\mathcal{L}_{\mathrm{str}}=
                    {\omega_{\mathrm{MYO}}\mathcal{L}_{\mathrm{MYO}}
                    +\omega_{\mathrm{LA}}\mathcal{L}_{\mathrm{LA}}}$
                    \State $\mathcal{L}\leftarrow\mathcal{L}
                    +{\lambda_{\mathrm{str}}}\mathcal{L}_{\mathrm{str}}$
                \EndIf
            \EndIf
            \State Update $\theta$ using Adam on $\mathcal{L}$
        \EndFor
        \State Evaluate validation performance on previously seen tasks
        \If{$t>1$}
            \State Compute instantaneous forgetting:
            \State $F^{(e)}=
            \frac{1}{t-1}\sum_{k<t}
            \max\left(0,S_k^{\mathrm{best}}-S_k^{(e)}\right)$
            \State Update EMA forgetting:
            \State $\bar{F}^{(e)}=
            \rho\bar{F}^{(e-1)}+(1-\rho)F^{(e)}$
            \If{$e>E_w$}
                \State Update replay coefficient for the next epoch:
                \State $\lambda_r^{(e+1)}=
                \operatorname{clip}\left(
                \lambda_r+\kappa\bar{F}^{(e)},
                \lambda_r^{\min},\lambda_r^{\max}\right)$
            \Else
                \State $\lambda_r^{(e+1)}\leftarrow\lambda_r$
            \EndIf
        \Else
            \State $\lambda_r^{(e+1)}\leftarrow\lambda_r$
        \EndIf
        \State Update $S_k^{\mathrm{best}}$ for monitored previous tasks
        \If{early stopping is satisfied}
            \State \textbf{break}
        \EndIf
    \EndFor
    \State Update $\mathcal{M}$ using
    Algorithm~\ref{alg:rarnet_memory}
\EndFor
\State \Return $f_{\theta}$
\end{algorithmic}
\end{algorithm}

Algorithm~\ref{alg:rarnet_training} summarizes the complete RAR-Net training procedure. Each task is learned for at most 20 epochs with early-stopping patience of eight epochs. Current-task mini-batches contain 16 samples and replay mini-batches contain eight historical examples. Mixed precision is used during training.

\begin{algorithm}[!t]
\caption{Protected Task-Aware Replay Memory Update.}
\label{alg:rarnet_memory}
\small
\begin{algorithmic}[1]

\Require Current-task data $\mathcal{D}_t$,
memory $\mathcal{M}$,
capacity $M$,
task-allocation weights $\{\pi_k\}$
\Ensure Updated replay memory $\mathcal{M}$

\For{each sample $(x_i,y_i)\in\mathcal{D}_t$}

    \State Compute class reliability $R_i^{c}$
    \State Compute boundary reliability $R_i^{b}$
    \State Compute image reliability $R_i^{I}$
    \State Compute anatomical reliability $R_i^{a}$
    \State Compute temporal reliability $R_i^{t}$

    \State Compute reliability-based replay priority:
    \Statex \hspace{\algorithmicindent}
    $\displaystyle
    p_i =
    \alpha R_i^{c}
    + \beta R_i^{b}
    + \gamma R_i^{I}
    + \delta R_i^{a}
    + \epsilon R_i^{t}
    $

    \State Determine the observed task set
    $\mathcal{T}_{\mathrm{seen}}$

    \For{each $k \in \mathcal{T}_{\mathrm{seen}}$}
        \State Compute dynamically normalized protected quota:
        \Statex \hspace{\algorithmicindent}
        $\displaystyle
        q_k =
        \operatorname{round}
        \left(
        M
        \frac{\pi_k}
        {\sum_{j\in\mathcal{T}_{\mathrm{seen}}}\pi_j}
        \right)
        $
    \EndFor

    \State Adjust integer quotas such that
    $\displaystyle
    \sum_{k\in\mathcal{T}_{\mathrm{seen}}} q_k = M$

    \If{$|\mathcal{M}| < M$}
        \State Insert $(x_i,y_i,p_i,t)$ into $\mathcal{M}$

    \ElsIf{task $t$ contains fewer than $q_t$ stored samples}
        \State Identify an overrepresented task $k$
        \State Find its lowest-priority exemplar $j$
        \If{$p_i > p_j$}
            \State Replace exemplar $j$ with $(x_i,y_i,p_i,t)$
        \EndIf

    \Else
        \State Find the lowest-priority exemplar $j$ belonging to task $t$
        \If{$p_i > p_j$}
            \State Replace exemplar $j$ with $(x_i,y_i,p_i,t)$
        \EndIf
    \EndIf

\EndFor

\State \Return $\mathcal{M}$

\end{algorithmic}
\end{algorithm}

At each optimization step, the current-task loss is first evaluated. For $t>1$, historical examples are sampled from protected memory and weighted by the current adaptive replay coefficient. Fully annotated historical samples additionally contribute the MYO-LA structure-aware objective. At the end of each epoch, all previously encountered validation tasks are evaluated, forgetting is estimated, and the EMA controller updates the replay coefficient. After task convergence, Algorithm~\ref{alg:rarnet_memory} updates the protected replay memory using the five-component reliability priority.

\section{Experimental Setup}
\label{experimental}

We designed the experiments to examine two related questions: how accurately the model segments cardiac structures as it encounters new echocardiographic domains, and how much previously acquired knowledge survives this sequential learning process. The evaluation therefore goes beyond final-task accuracy. Three heterogeneous echocardiographic datasets were arranged as a sequential learning problem, and each model was evaluated after every learning stage. This allowed segmentation quality and CL behavior to be examined together. For a controlled comparison, RAR-Net and all baseline methods used the same backbone, data partitions, training schedule, and core optimization settings. Experiments were repeated using three deterministic random seeds (42, 123, and 2025), and the resulting performance is reported as mean $\pm$ standard deviation unless stated otherwise.

\subsection{Datasets}
\label{sec:datasets}

The experiments were conducted on three echocardiographic datasets: CAMUS, CardiacNet, and EchoNet-Dynamic. These datasets differ not only in their acquisition characteristics and patient populations, but also in the type of anatomical supervision they provide. CAMUS and CardiacNet contain multi-structure cardiac annotations, whereas EchoNet-Dynamic provides left-ventricular contour annotations. This difference makes the resulting CL problem more realistic: the model encounters both a change in image distribution and a change in available supervision as training progresses.

\subsubsection{CAMUS}

The CAMUS dataset \cite{leclerc2019deep} contains two-dimensional echocardiographic examinations from 500 patients. Each examination includes apical two-chamber (2CH) and four-chamber (4CH) views, with expert annotations provided at end-diastole (ED) and end-systole (ES). The dataset contains patients with different image qualities and cardiac conditions, making it a useful benchmark for multi-structure echocardiographic segmentation. We followed the official patient-level partition: 400 patients were used for training, 50 for validation, and the remaining 50 for testing. Keeping the partition at the patient level ensures that images originating from the same patient cannot appear across training and evaluation subsets.

For our experiments, the segmentation label space was standardized to four classes: BG, LV, MYO, and LA, represented by labels 0, 1, 2, and 3, respectively. Only the annotated ED and ES acquisitions were used. Echocardiographic images and their corresponding masks were resized to $256\times256$ pixels. Image intensities were normalized to the range $[0,1]$, while nearest-neighbor interpolation was used for segmentation masks to preserve discrete anatomical labels.

\subsubsection{CardiacNet}

CardiacNet \cite{yang2024cardiacnet} provides echocardiographic data collected for cardiac disease assessment and introduces a clinically different domain from CAMUS. This domain shift is useful in our setting because continual segmentation should remain effective when the acquisition distribution and patient population change rather than assuming that successive tasks are nearly identical. To maintain consistency with CAMUS, the CardiacNet annotations were mapped to the common four-class label space comprising BG, LV, MYO, and LA. The additional label 4 present in the original annotations was mapped to background. Slices containing no foreground cardiac structure after label mapping were excluded. CardiacNet was randomly divided at the patient level using the same proportion as CAMUS: 80\% of the patients were assigned to training, 10\% to validation, and 10\% to testing. The patient-level constraint was applied before individual two-dimensional samples were generated. Consequently, frames or slices belonging to a given patient remained within a single partition throughout the experiments. The same preprocessing used for CAMUS was applied to CardiacNet. Images and segmentation masks were resized to $256\times256$ pixels, and image intensities were normalized to $[0,1]$.

\subsubsection{EchoNet-Dynamic}

EchoNet-Dynamic \cite{ouyang2020video} is a large-scale echocardiographic video dataset containing 10,030 apical four-chamber videos acquired during routine clinical practice. In contrast to CAMUS and CardiacNet, EchoNet-Dynamic provides expert LV tracings rather than complete LV-MYO-LA segmentation annotations.  Expert contour coordinates were converted into binary LV masks, and the corresponding annotated frames were extracted from the original videos. Frames and masks were subsequently resized to $256\times256$ pixels. Because MYO and LA GT annotations are unavailable in EchoNet-Dynamic, these structures were not treated as background during optimization. Instead, EchoNet-Dynamic samples were handled using LV-only partial supervision. Evaluation on this dataset was likewise restricted to the LV. This distinction is particularly relevant to the CL setting: the final task changes not only the image domain but also the available anatomical supervision. For EchoNet-Dynamic, we follow the original dataset partition.

\begin{figure*}[!t] 
\centering 
\includegraphics[width=1\textwidth]{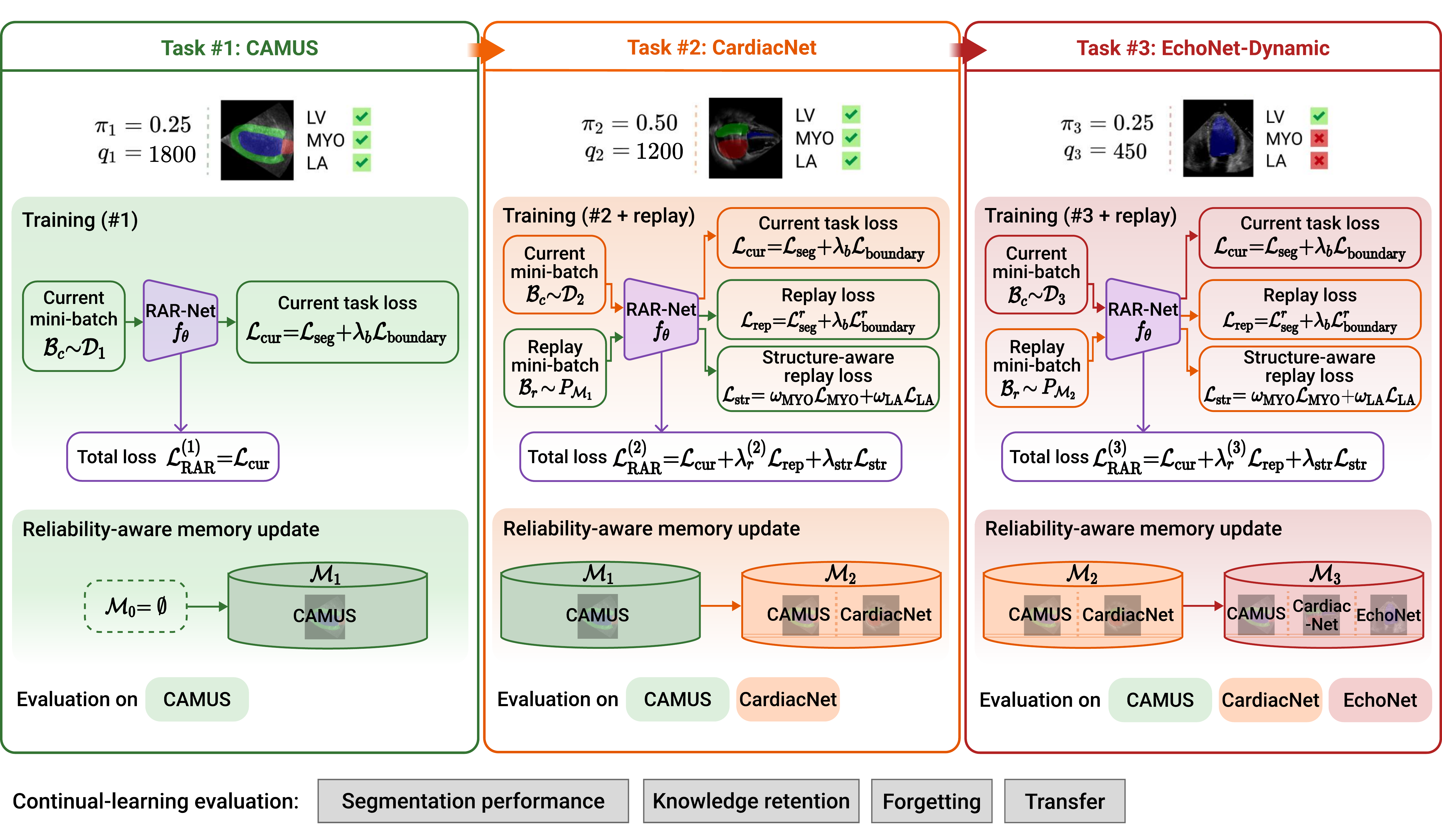} 
\caption{CL evaluation protocol of RAR-Net across CAMUS, CardiacNet, and EchoNet-Dynamic. The first two datasets provide LV, MYO, and LA annotations, whereas EchoNet-Dynamic provides LV-only supervision. From the second task onward, training incorporates protected replay memory with reliability-aware sample selection and stage-wise normalized task-allocation weights. After each task, the model is evaluated on all previously observed domains to assess segmentation performance, knowledge retention, forgetting, and transfer.} 
\label{protocol} 
\end{figure*}

\subsection{Experimental Protocol}
\label{sec:experimental_protocol}

{Figure~\ref{protocol} summarizes the sequential training, replay-memory update, and evaluation protocol used across the three tasks.}
The primary CL sequence was defined as

\begin{equation}
\begin{aligned}
\mathcal{T}_1:\mathrm{CAMUS}
&\rightarrow
\mathcal{T}_2:\mathrm{CardiacNet}
&\rightarrow
\mathcal{T}_3:\mathrm{EchoNet\text{-}Dynamic}.
\end{aligned}
\label{eq:task_order}
\end{equation}

A single segmentation model was trained sequentially according to this task order. After completing a task, training proceeded to the next dataset without unrestricted access to the complete training data of previous tasks. For rehearsal-based methods, historical information could be accessed only through the predefined replay memory. This restriction separates the CL methods from joint training, which retains simultaneous access to all training domains.

A common replay-memory capacity of 1,800 samples was used for RAR-Net and the replay-based baselines. The current-task batch size was 16, while eight replay samples were drawn for each replay mini-batch. These settings were kept fixed across replay-based methods so that differences in performance were not simply caused by giving one method a larger memory or greater replay budget. After each task, the current model was evaluated on the available evaluation sets of all three datasets. The resulting scores were organized into an accuracy matrix

\begin{equation}
\mathbf{A}\in\mathbb{R}^{T\times T},
\end{equation}

where $A_{i,j}$ denotes the segmentation performance on task $j$ after the model has completed training on task $i$. For example, $A_{3,1}$ measures performance on CAMUS after the model has subsequently learned CardiacNet and EchoNet-Dynamic. The difference between this value and the earlier CAMUS performance provides a direct indication of whether the original anatomical knowledge was retained or forgotten. All experiments were repeated using random seeds 42, 123, and 2025. The same dataset partitions were maintained across compared methods for a given experimental run. Random number generation was controlled for Python, NumPy, and PyTorch, and deterministic CUDA behavior was enabled where supported. This multi-seed evaluation was used to reduce the possibility that conclusions were driven by a favorable initialization or stochastic training trajectory.

\subsection{Implementation Details}
\label{sec:implementation_details}

All models were implemented in PyTorch and trained using CUDA acceleration. The segmentation backbone follows a U-Net-style encoder-decoder architecture with skip connections and a base channel width of 32. A single-channel $256\times256$ echocardiographic image is provided as input, and the segmentation branch predicts four output classes corresponding to BG, LV, MYO, and LA. A parallel boundary-prediction branch is attached to the final decoder representation for methods that make use of boundary information.

The Adam optimizer was used with an initial learning rate of $1\times10^{-4}$ and weight decay of $1\times10^{-5}$. Each sequential task was trained for a maximum of 20 epochs. Joint training was likewise performed for 20 epochs to maintain a comparable optimization budget. Early stopping was triggered when validation performance failed to improve for eight consecutive epochs.  The training batch size was 16 and the replay batch size was 8. Mixed-precision training was enabled to reduce GPU-memory consumption and improve computational efficiency. The replay loss was assigned a base weight of 1.00, while the boundary component was weighted by 0.30. The replay memory capacity was fixed at 1,800 samples. These core settings were shared by the proposed model and the corresponding baselines wherever applicable. Table \ref{tab:training_hyperparameters} summarizes the common training configuration.

\begin{table}[t]
\centering
\caption{Training hyperparameters used in the experiments.}
\label{tab:training_hyperparameters}
\begin{tabular}{ll}
\toprule
\textbf{Hyperparameter} & \textbf{Setting} \\
\midrule
Input image size          & $256\times256$ \\
Number of classes         & 4 \\
{Base channel width} & 32 \\
Optimizer                  & Adam \\
Learning rate              & $1\times10^{-4}$ \\
Weight decay               & $1\times10^{-5}$ \\
Batch size                 & 16 \\
Replay batch size          & 8 \\
Training epochs            & 20 \\
Joint-training epochs      & 20 \\
Early-stopping patience    & 8 epochs \\
Mixed-precision training   & Enabled \\
Replay memory capacity     & {1,800} \\
{Boundary loss coefficient} & 0.30 \\
{Base replay coefficient} & 1.00 \\
Random seeds               & 42, 123, 2025 \\
\bottomrule
\end{tabular}
\end{table}

\subsection{Evaluation Metrics}
\label{sec:evaluation_metrics}

Continual segmentation cannot be judged reliably from final Dice alone. A model may perform well on EchoNet-Dynamic after the last training stage while having lost much of its earlier CAMUS or CardiacNet knowledge. We therefore evaluate the models from three perspectives: CL behavior, segmentation quality, and computational efficiency.

\subsubsection{Continual-Learning Metrics}
\label{sec:cl_metrics}

Following established CL evaluation protocols \cite{lopez2017gradient,chaudhry2018riemannian}, four measures were derived from the task-performance matrix $\mathbf{A}$: Final Average Accuracy (FAA), Average Forgetting (AF), Backward Transfer (BWT), and Forward Transfer (FWT).

\paragraph{Final Average Accuracy.}
FAA measures the average segmentation performance retained across all $T$ tasks after completing the entire learning sequence:

\begin{equation}
\mathrm{FAA}
=
\frac{1}{T}
\sum_{j=1}^{T}A_{T,j}.
\label{eq:faa}
\end{equation}

A larger FAA indicates a more favorable overall balance between acquiring new knowledge and retaining earlier knowledge.

\paragraph{Average Forgetting.}
AF measures the change in performance on previously learned tasks after subsequent training. It is computed as

\begin{equation}
\mathrm{AF}
=
\frac{1}{T-1}
\sum_{j=1}^{T-1}
\left[
\max_{i\in\{j,\ldots,T-1\}}A_{i,j}
-
A_{T,j}
\right].
\label{eq:af}
\end{equation}

Here, $A_{i,j}$ denotes the Dice score on task $j$ after training through task $i$, and $T$ represents the total number of tasks. The historical maximum is calculated over evaluations performed after task $j$ has been learned and before the final training stage. AF is reported as a signed quantity without clipping negative values to zero. Positive AF indicates a net decline relative to the best historical performance, whereas negative AF indicates that the final model achieves higher performance than the corresponding historical reference on average. Lower AF is preferable, with zero indicating no net change relative to the historical maximum. Under the adopted maximum-based definition, AF may be negative when the final performance on previously
learned tasks exceeds their best recorded performance before the final task. Negative AF therefore indicates net improvement rather than forgetting. In the present experiments, the maximum historical performance for each previous task coincides with its performance immediately after training, yielding AF = -BWT for the evaluated task sequences.

\paragraph{Backward Transfer.}
BWT measures the effect of subsequent learning on tasks that have already been learned:

\begin{equation}
\mathrm{BWT}
=
\frac{1}{T-1}
\sum_{j=1}^{T-1}
\left(
A_{T,j}-A_{j,j}
\right).
\label{eq:bwt}
\end{equation}

Negative BWT indicates interference with earlier tasks, whereas positive BWT suggests that later learning improves representations useful for previously encountered domains.

\paragraph{Forward Transfer.}
FWT evaluates how useful knowledge acquired from earlier tasks is for subsequent domains before those domains are explicitly learned. In our implementation, FWT is obtained from the available upper-triangular elements
of $\mathbf{A}$:

\begin{equation}
\mathrm{FWT}
=
\frac{1}{|\mathcal{U}|}
\sum_{(i,j)\in\mathcal{U}}A_{i,j},
\qquad
\mathcal{U}=\{(i,j)\mid i<j\}.
\label{eq:fwt}
\end{equation}

Under this definition, FWT represents absolute pre-exposure performance rather than improvement relative to a randomly initialized reference model. Higher values therefore indicate greater transferability of representations learned from earlier echocardiographic domains.

\subsubsection{Segmentation Metrics}
\label{sec:segmentation_metrics}

Segmentation quality was evaluated using the Dice similarity coefficient, intersection over union (IoU), 95th-percentile Hausdorff distance (HD95), average symmetric surface distance (ASD), and Boundary F1 (BF1). Together, these metrics capture both region-level agreement and anatomical boundary quality. For predicted foreground region $P$ and GT region $G$, Dice is

\begin{equation}
\mathrm{Dice}
=
\frac{2|P\cap G|}
{|P|+|G|},
\label{eq:dice}
\end{equation}

and IoU is

\begin{equation}
\mathrm{IoU}
=
\frac{|P\cap G|}
{|P\cup G|}.
\label{eq:iou}
\end{equation}

Both range from 0 to 1, with larger values indicating greater overlap.

Overlap alone, however, does not fully characterize anatomical segmentation quality. Two predictions may obtain similar Dice scores while differing substantially in contour localization. We therefore additionally report HD95 and ASD \cite{taha2015metrics}. HD95 measures the 95th percentile of the bidirectional distances between predicted and reference surfaces, reducing the sensitivity of the conventional Hausdorff distance to isolated extreme outliers. ASD measures the mean symmetric distance between the two surfaces. For both metrics, lower values indicate better boundary agreement. BF1 provides a complementary contour-oriented measure. Let $P_b$ and $R_b$ denote boundary precision and boundary recall, respectively. Boundary F1 is calculated as

\begin{equation}
\mathrm{Boundary~F1}
=
\frac{2P_bR_b}{P_b+R_b}.
\label{eq:boundary_f1}
\end{equation}

A two-pixel tolerance was used when determining whether predicted and GT boundary pixels correspond. Higher BF1 therefore indicates more accurate localization of cardiac contours. For CAMUS and CardiacNet, class-wise metrics were computed for LV, MYO, and LA and subsequently summarized across the foreground structures. Because EchoNet-Dynamic provides only LV contour annotations, its segmentation evaluation was restricted to LV.

\subsubsection{Computational Metrics}
\label{sec:computational_metrics}

CL improvements are less convincing if they require a disproportionate increase in computation. We therefore report four computational measures: number of trainable parameters, training time, peak GPU-memory consumption, and evaluation runtime per image. The number of trainable parameters was obtained directly from each model. Training time was measured as the elapsed wall-clock time required for each task-training stage. Peak GPU memory was recorded using CUDA memory statistics during optimization. Evaluation runtime per image was calculated from the elapsed evaluation time normalized by the number of processed images:

\begin{equation}
{t_{\mathrm{eval/img}}}
=
\frac{t_{\mathrm{eval}}}{N},
\label{eq:inference_time}
\end{equation}

where $t_{\mathrm{eval}}$ is the total evaluation time and $N$ is the number of evaluated images. These measurements allow the accuracy gains of RAR-Net to be considered alongside its practical computational cost.

\begin{table}[!htbp]
\centering
\scriptsize
\caption{Computational characteristics of the proposed RAR-Net 
across the sequential learning tasks. Training time is reported as 
mean $\pm$ standard deviation over three random seeds. 
{EchoNet denotes EchoNet-Dynamic.}}
\label{tab:rarnet_computation}

\small
\setlength{\tabcolsep}{3pt}
\renewcommand{\arraystretch}{1.12}

\begin{tabularx}{\columnwidth}{
@{}
>{\raggedright\arraybackslash}p{2.2cm}
>{\centering\arraybackslash}X
>{\centering\arraybackslash}p{2.2cm}
@{}
}
\toprule
\textbf{Task} &
\textbf{Training time (s)} &
\makecell{\textbf{Peak GPU} \\ \textbf{memory (GB)}} \\
\midrule

CAMUS &
$786.95 \pm 10.02$ &
1.6620 \\

CardiacNet &
{$16{,}802.10 \pm 65.44$} &
2.3114 \\

{EchoNet} &
{$6{,}765.76 \pm 16.07$} &
2.2710 \\

\bottomrule
\end{tabularx}

\vspace{2pt}
\begin{minipage}{\columnwidth}
\footnotesize
{The model contains 7.76 M trainable parameters, which remain unchanged across all sequential learning tasks.}
\end{minipage}

\end{table}

Table~\ref{tab:rarnet_computation} summarizes the computational characteristics of RAR-Net across the sequential learning stages. The parameter count remains fixed at 7.76 M throughout the sequence, indicating that continual adaptation does not require expansion of the segmentation network. The peak GPU memory usage ranges from 1.6620 GB for CAMUS to 2.3114 GB for CardiacNet. Training time varies across tasks, with CardiacNet requiring the longest optimization time, followed by EchoNet-Dynamic and CAMUS.

\begin{table}[!htbp]
\centering
\caption{Method-specific hyperparameters used for the CL
baselines and the proposed RAR-Net.}
\label{tab:method_specific_hyperparameters}

\small
\setlength{\tabcolsep}{5pt}
\renewcommand{\arraystretch}{1.10}

\begin{tabularx}{\textwidth}{
@{}
>{\raggedright\arraybackslash}p{3.2cm}
>{\raggedright\arraybackslash}p{3.0cm}
>{\raggedright\arraybackslash}X
>{\centering\arraybackslash}p{2.0cm}
@{}
}
\toprule
\textbf{Method} &
\textbf{Component} &
\textbf{Hyperparameter} &
\textbf{Value} \\
\midrule

\multirow{2}{=}{Boundary-Balanced Replay}
& \multirow{2}{=}{Sampling}
& Class sampling weight ($\alpha_s$) & 0.40 \\
&
& Boundary sampling weight ($\beta_s$) & 0.60 \\
\addlinespace[2pt]

EWC
& Regularization
& Regularization coefficient ($\lambda_{\mathrm{EWC}}$) & 20.0 \\
\addlinespace[2pt]

\multirow{2}{=}{LwF}
& \multirow{2}{=}{Distillation}
& Distillation coefficient ($\lambda_{\mathrm{LwF}}$) & 0.50 \\
&
& Distillation temperature ($T$) & 2.0 \\
\addlinespace[2pt]

Plain Replay
& Replay
& Replay loss coefficient ($\lambda_r$) & 1.00 \\
\addlinespace[2pt]

iCaRL-style Replay
& Replay
& Replay loss coefficient ($\lambda_r$) & 1.00 \\
\addlinespace[2pt]

\multirow{2}{=}{{Weighted-Sampler Replay}}
& \multirow{2}{=}{Sampling}
& Class sampling weight ($\alpha_s$) & 0.40 \\
&
& Boundary sampling weight ($\beta_s$) & 0.60 \\

\midrule

\multirow[t]{19}{=}{\textbf{RAR-Net (Ours)}}

& \multirow[t]{5}{=}{Memory configuration}
& Replay memory capacity ($M$) & {1,800} \\
&
& Replay mini-batch size ($B_r$) & 8 \\
&
& {CAMUS allocation weight ($\pi_{\mathrm{CAMUS}}$)} & 0.25 \\
&
& {CardiacNet allocation weight ($\pi_{\mathrm{CardiacNet}}$)} & 0.50 \\
&
& {EchoNet-Dynamic allocation weight ($\pi_{\mathrm{EchoNet}}$)} & 0.25 \\
\cmidrule(lr){2-4}

&
\multirow[t]{6}{=}{Adaptive replay}
& Base replay coefficient ($\lambda_r$) & 1.00 \\
&
& Adaptive replay minimum ($\lambda_r^{\min}$) & 0.75 \\
&
& Adaptive replay maximum ($\lambda_r^{\max}$) & 1.75 \\
&
& Adaptive replay gain ($\kappa$) & 4.00 \\
&
& Forgetting EMA coefficient ($\rho$) & 0.70 \\
&
& Adaptive replay warm-up & 3 epochs \\
\cmidrule(lr){2-4}

&
\multirow[t]{5}{=}{Reliability scoring}
& Class reliability weight ($\alpha$) & 0.15 \\
&
& Boundary reliability weight ($\beta$) & 0.20 \\
&
& Image reliability weight ($\gamma$) & 0.20 \\
&
& Anatomical reliability weight ($\delta$) & 0.25 \\
&
& Temporal reliability weight ($\epsilon$) & 0.20 \\
\cmidrule(lr){2-4}

&
\multirow[t]{3}{=}{Structure-aware replay}
& Structure-aware replay coefficient ($\lambda_{\mathrm{str}}$) & 0.30 \\
&
& MYO/LA replay weights
  ({$\omega_{\mathrm{MYO}}, \omega_{\mathrm{LA}}$})
& $(1.25,\;1.00)$ \\
&
& Tversky parameters ($\alpha_T,\beta_T$)
& $(0.30,\;0.70)$ \\

\bottomrule
\end{tabularx}
\end{table}

For a consistent comparison, all sequential methods were trained using the same backbone, data preprocessing, optimization settings, and task sequence, while method-specific hyperparameters were configured according to their respective formulations. The complete baseline-specific settings and the hyperparameters used for the proposed RAR-Net are summarized in Table~\ref{tab:method_specific_hyperparameters}.

\section{Results}
\label{sec:results}

This section examines the performance of RAR-Net under sequential training on CAMUS, CardiacNet, and EchoNet-Dynamic. We first compare overall performance against established CL methods and the offline joint-training reference. We then examine knowledge retention, transfer, and dataset-specific segmentation quality, with particular attention to the trade-offs introduced by heterogeneous domains and annotation protocols. All results are reported as mean $\pm$ standard deviation over three random seeds (42, 123, and 2025).

\subsection{Overall Comparison}
\label{sec:overall_results}

Table~\ref{tab:overall_comparison} compares RAR-Net with established CL approaches, including regularization-based methods, conventional replay, DER, and MIR. RAR-Net achieves the highest FAA among the evaluated CL methods, reaching $0.8718\pm0.0029$. The strongest competing method by FAA is Weighted-Sampler Replay ($0.8636\pm0.0047$), followed by Plain Replay ($0.8544\pm0.0065$) and MIR ($0.8534\pm0.0048$). Although these differences are relatively modest, they suggest that reliability-guided memory management provides an advantage over conventional rehearsal under the evaluated task sequence.

The offline Joint Training reference achieves an FAA of $0.8726\pm0.0140$, only $0.0008$ above RAR-Net. Unlike CL methods, Joint Training has simultaneous access to the complete training data from all three datasets. Thus, the comparison indicates that RAR-Net approaches the performance of offline optimization while operating with bounded historical replay memory. However, the small numerical difference should not be interpreted as evidence of statistical equivalence.

The baseline results also reveal substantial variation in the ability to retain knowledge across domains. EWC and LwF perform poorly overall, despite their mechanisms for limiting changes to previously learned representations or outputs. Similarly, Boundary-Balanced Replay retains relatively strong performance on CAMUS but struggles on CardiacNet. In contrast, RAR-Net maintains comparatively balanced segmentation performance across the three datasets. This distinction is relevant because high accuracy on the most recently learned domain does not necessarily indicate successful CL.

\subsection{Continual-Learning and Dataset-Wise Segmentation Performance}
\label{sec:cl_results}
\label{sec:dataset_results}

The CL metrics in Table~\ref{tab:overall_comparison}(a) clarify the extent to which the overall improvements arise from historical knowledge retention. RAR-Net achieves an AF value of $-0.0489\pm0.0116$ and BWT of $0.0489\pm0.0116$, outperforming the corresponding retention metrics of Plain Replay and Weighted-Sampler Replay. Under the maximum-based forgetting definition adopted in this study, negative AF indicates that final performance on previously learned tasks exceeds their best recorded performance before the final task. The positive BWT is consistent with this interpretation, suggesting that subsequent training can improve previously acquired segmentation performance rather than merely preventing its deterioration.

\begin{table}[!t]
\centering
\caption{Comparison of CL methods after sequential
training on CAMUS $\rightarrow$ CardiacNet $\rightarrow$
EchoNet-Dynamic. Results are reported as mean $\pm$ standard
deviation over three random seeds (42, 123, and 2025).
Higher FAA, BWT, FWT, and Dice indicate better performance,
whereas lower AF indicates better performance.
Joint Training is included as an offline full-data reference;
therefore, AF, BWT, and FWT are not applicable.
Boldface indicates the best mean among CL methods.}
\label{tab:overall_comparison}

\footnotesize
\setlength{\tabcolsep}{3pt}
\renewcommand{\arraystretch}{1.15}


\begin{tabularx}{\textwidth}{@{}
>{\raggedright\arraybackslash}p{3.0cm}
*{4}{>{\centering\arraybackslash}X}
@{}}

\toprule
\multicolumn{5}{@{}l}{
\textit{(a) Continual-learning metrics}} \\
\midrule

\textbf{Method}
& \textbf{FAA $\uparrow$}
& \textbf{AF $\downarrow$}
& \textbf{BWT $\uparrow$}
& \textbf{FWT $\uparrow$} \\
\midrule

Joint Training (Offline)
& $0.8726 \pm 0.0140$
& N/A
& N/A
& N/A \\

Plain Replay
& $0.8544 \pm 0.0065$
& $-0.0318 \pm 0.0246$
& $0.0318 \pm 0.0246$
& $0.4269 \pm 0.0340$ \\

Boundary-Balanced Replay
& $0.7047 \pm 0.0167$
& $0.1919 \pm 0.0079$
& $-0.1919 \pm 0.0079$
& $0.4334 \pm 0.0229$ \\

EWC
& $0.5083 \pm 0.0055$
& $0.5552 \pm 0.0087$
& $-0.5552 \pm 0.0087$
& $0.3535 \pm 0.0800$ \\

LwF
& $0.4438 \pm 0.0256$
& $0.0869 \pm 0.0347$
& $-0.0858 \pm 0.0344$
& $0.3843 \pm 0.0097$ \\

iCaRL-style Replay
& $0.6366 \pm 0.0173$
& $0.2947 \pm 0.0192$
& $-0.2947 \pm 0.0192$
& $0.4334 \pm 0.0229$ \\

Weighted-Sampler Replay
& $0.8636 \pm 0.0047$
& $-0.0254 \pm 0.0090$
& $0.0254 \pm 0.0090$
& $0.3950 \pm 0.0423$ \\

DER
& $0.7934 \pm 0.0378$
& $0.0282 \pm 0.0018$
& $-0.0282 \pm 0.0018$
& $0.4173 \pm 0.0070$ \\

MIR
& $0.8534 \pm 0.0048$
& $-0.0243 \pm 0.0267$
& $0.0243 \pm 0.0267$
& $0.3998 \pm 0.0428$ \\

\midrule

\textbf{RAR-Net (Ours)}
& $\mathbf{0.8718 \pm 0.0029}$
& $\mathbf{-0.0489 \pm 0.0116}$
& $\mathbf{0.0489 \pm 0.0116}$
& $\mathbf{0.4366 \pm 0.0357}$ \\

\bottomrule
\end{tabularx}

\vspace{8pt}


\begin{tabularx}{\textwidth}{@{}
>{\raggedright\arraybackslash}p{3.0cm}
*{3}{>{\centering\arraybackslash}X}
@{}}

\toprule
\multicolumn{4}{@{}l}{
\textit{(b) Final Dice after sequential learning}} \\
\midrule

\textbf{Method}
& \textbf{CAMUS $\uparrow$}
& \textbf{CardiacNet $\uparrow$}
& \textbf{EchoNet-Dynamic $\uparrow$} \\
\midrule

Joint Training (Offline)
& $0.8948 \pm 0.0059$
& $0.8196 \pm 0.0338$
& $0.9035 \pm 0.0031$ \\

Plain Replay
& $0.8735 \pm 0.0193$
& $0.7808 \pm 0.0067$
& $\mathbf{0.9090 \pm 0.0015}$ \\

Boundary-Balanced Replay
& $0.8935 \pm 0.0004$
& $0.3154 \pm 0.0483$
& $0.9053 \pm 0.0017$ \\

EWC
& $0.4214 \pm 0.0136$
& $0.1950 \pm 0.0314$
& $0.9085 \pm 0.0046$ \\

LwF
& $0.7285 \pm 0.0827$
& $0.1328 \pm 0.0053$
& $0.4702 \pm 0.0257$ \\

iCaRL-style Replay
& $0.8934 \pm 0.0011$
& $0.1099 \pm 0.0526$
& $0.9064 \pm 0.0022$ \\

Weighted-Sampler Replay
& $0.8841 \pm 0.0023$
& $0.7986 \pm 0.0138$
& $0.9081 \pm 0.0014$ \\

DER
& $0.8760 \pm 0.0017$
& $0.6021 \pm 0.1064$
& $0.9019 \pm 0.0072$ \\

MIR
& $0.8880 \pm 0.0059$
& $0.7646 \pm 0.0186$
& $0.9078 \pm 0.0014$ \\

\midrule

\textbf{RAR-Net (Ours)}
& $\mathbf{0.8977 \pm 0.0024}$
& $\mathbf{0.8096 \pm 0.0055}$
& $0.9081 \pm 0.0023$ \\

\bottomrule
\end{tabularx}
\end{table}

RAR-Net also obtains the highest reported forward-transfer measure ($0.4366\pm0.0357$), although the difference from Boundary-Balanced Replay and iCaRL-style Replay is small. Since this measure represents absolute pre-exposure performance rather than baseline-corrected FWT, it should be interpreted as an indication of cross-domain generalization before explicit training on the subsequent task. In summary, the results suggest that RAR-Net's main advantage lies in retaining and consolidating historical knowledge, rather than producing a substantial improvement in FWT.

The final dataset-specific Dice scores in Table~\ref{tab:overall_comparison}(b) help explain this behavior. RAR-Net achieves the highest Dice among CL methods on CAMUS ($0.8977\pm0.0024$) and CardiacNet ($0.8096\pm0.0055$). CAMUS is particularly informative because it is learned first and must remain accessible throughout two subsequent training stages. Its final performance is comparable to the offline Joint Training reference, indicating that the proposed replay mechanisms preserve much of the knowledge acquired during initial training.

CardiacNet presents a more demanding retention problem. As the intermediate task, it requires adaptation from CAMUS while remaining resistant to interference from the subsequent LV-only EchoNet-Dynamic dataset. RAR-Net maintains a CardiacNet Dice of $0.8096$, compared with $0.7986$ for Weighted-Sampler Replay and $0.7808$ for Plain Replay. The gap is considerably larger against Boundary-Balanced Replay and iCaRL-style Replay, both of which retain relatively strong CAMUS performance but deteriorate substantially on CardiacNet. This contrast suggests that protecting historical samples from multiple tasks is more useful than preserving the earliest task alone.

The picture is slightly different on EchoNet-Dynamic. RAR-Net achieves a Dice of $0.9081\pm0.0023$, whereas Plain Replay obtains $0.9090\pm0.0015$. The difference is small, but it shows that RAR-Net does not uniformly outperform conventional replay on the final task. Instead, its advantage emerges from maintaining stronger performance on the earlier datasets while preserving competitive adaptation to the incoming LV-only supervision. This is a more meaningful outcome for continual segmentation than maximizing final-task accuracy at the expense of previously learned structures.

Table~\ref{tab:additional_segmentation_metrics} provides further evidence through region-overlap and boundary-distance metrics. RAR-Net achieves the highest IoU and lowest HD95 and ASD among the evaluated CL methods across all three datasets. These improvements are particularly relevant on CAMUS and CardiacNet, where preservation of previously learned anatomical boundaries remains necessary after subsequent training. Nevertheless, the BF1 results are less uniform: Weighted-Sampler Replay performs slightly better on CardiacNet, while EWC achieves the highest BF1 on EchoNet-Dynamic. Thus, the improvements in region overlap and surface distance do not translate into superiority on every boundary metric.

Overall, the results indicate that RAR-Net provides a favorable balance between adaptation and historical knowledge retention under the evaluated task sequence. Its strongest gains occur on previously learned datasets, particularly CardiacNet, while performance on the final EchoNet-Dynamic task remains competitive. This pattern is consistent with the objective of reliability-guided replay: retaining useful historical information without substantially restricting adaptation to newly introduced data.

\begin{table*}[!t]
\centering
\caption{Comparison of final segmentation performance after
sequential learning on CAMUS $\rightarrow$ CardiacNet
$\rightarrow$ EchoNet-Dynamic. Results are reported as mean
$\pm$ standard deviation over three random seeds (42, 123,
and 2025). Higher IoU and BF1 indicate better
performance, whereas lower HD95 and ASD indicate better
performance. Joint Training is included as an offline
full-data reference and is excluded from the boldface
comparison. Dice is reported separately in
Table~\ref{tab:overall_comparison}.}
\label{tab:additional_segmentation_metrics}

\footnotesize
\setlength{\tabcolsep}{3pt}
\renewcommand{\arraystretch}{1.12}

\begin{tabularx}{\textwidth}{@{}
>{\raggedright\arraybackslash}p{3.0cm}
*{4}{>{\centering\arraybackslash}X}
@{}}

\toprule
\multicolumn{5}{@{}l}{\textit{(a) CAMUS}} \\
\midrule
\textbf{Method}
& \textbf{IoU $\uparrow$}
& \textbf{HD95 $\downarrow$}
& \textbf{ASD $\downarrow$}
& \textbf{BF1 $\uparrow$} \\
\midrule

Joint Training (Offline)
& $0.8112 \pm 0.0097$
& $8.3562 \pm 1.2227$
& $2.9143 \pm 0.2741$
& $0.5748 \pm 0.0155$ \\

Plain Replay
& $0.7782 \pm 0.0287$
& $9.6623 \pm 0.7356$
& $3.4373 \pm 0.3775$
& $0.5311 \pm 0.0251$ \\

Boundary-Balanced Replay
& $0.8090 \pm 0.0006$
& $8.6206 \pm 0.5460$
& $3.0078 \pm 0.0850$
& $0.5736 \pm 0.0066$ \\

EWC
& $0.3268 \pm 0.0206$
& $43.5470 \pm 3.5978$
& $14.2110 \pm 1.5009$
& $0.1527 \pm 0.0217$ \\

LwF
& $0.5921 \pm 0.0968$
& $18.0455 \pm 4.0371$
& $6.8726 \pm 1.5564$
& $0.3618 \pm 0.0847$ \\

iCaRL-style Replay
& $0.8088 \pm 0.0019$
& $8.8615 \pm 0.5892$
& $3.0611 \pm 0.1052$
& $0.5681 \pm 0.0070$ \\

Weighted-Sampler Replay
& $0.7941 \pm 0.0038$
& $9.1843 \pm 0.3616$
& $3.2485 \pm 0.0798$
& $0.5490 \pm 0.0062$ \\

DER
& $0.7814 \pm 0.0030$
& $21.1039 \pm 21.3896$
& $5.6569 \pm 4.1085$
& $0.5069 \pm 0.0268$ \\

MIR
& $0.8002 \pm 0.0095$
& $8.6204 \pm 0.3489$
& $3.0953 \pm 0.1001$
& $0.5525 \pm 0.0078$ \\

\midrule

\textbf{RAR-Net (Ours)}
& $\mathbf{0.8114 \pm 0.0044}$
& $\mathbf{8.2862 \pm 0.5435}$
& $\mathbf{2.9359 \pm 0.1164}$
& $\mathbf{0.5750 \pm 0.0096}$ \\

\midrule
\multicolumn{5}{@{}l}{\textit{(b) CardiacNet}} \\
\midrule
\textbf{Method}
& \textbf{IoU $\uparrow$}
& \textbf{HD95 $\downarrow$}
& \textbf{ASD $\downarrow$}
& \textbf{BF1 $\uparrow$} \\
\midrule

Joint Training (Offline)
& $0.6953 \pm 0.0479$
& $9.2428 \pm 3.9679$
& $3.5307 \pm 1.0161$
& $0.6131 \pm 0.0376$ \\

Plain Replay
& $0.6411 \pm 0.0092$
& $16.4699 \pm 1.3680$
& $5.6070 \pm 0.5448$
& $0.5568 \pm 0.0113$ \\

Boundary-Balanced Replay
& $0.2013 \pm 0.0369$
& $56.9508 \pm 2.7925$
& $25.3373 \pm 1.2264$
& $0.1476 \pm 0.0197$ \\

EWC
& $0.1262 \pm 0.0265$
& $80.8526 \pm 5.0313$
& $37.5184 \pm 1.7394$
& $0.1346 \pm 0.0214$ \\

LwF
& $0.0772 \pm 0.0046$
& $62.6021 \pm 2.9053$
& $28.6867 \pm 1.0276$
& $0.0971 \pm 0.0073$ \\

iCaRL-style Replay
& $0.0633 \pm 0.0333$
& $61.4584 \pm 1.0873$
& $36.9449 \pm 2.0463$
& $0.0970 \pm 0.0008$ \\

Weighted-Sampler Replay
& $0.6658 \pm 0.0195$
& $18.2134 \pm 1.1984$
& $5.9187 \pm 0.2670$
& $\mathbf{0.5831 \pm 0.0219}$ \\

DER
& $0.4447 \pm 0.1071$
& $49.6329 \pm 27.5366$
& $19.0059 \pm 11.6101$
& $0.3144 \pm 0.0887$ \\

MIR
& $0.6206 \pm 0.0239$
& $18.8856 \pm 4.3442$
& $6.5883 \pm 1.3403$
& $0.5252 \pm 0.0324$ \\

\midrule

\textbf{RAR-Net (Ours)}
& $\mathbf{0.6777 \pm 0.0106}$
& $\mathbf{16.0261 \pm 0.9857}$
& $\mathbf{5.2089 \pm 0.1841}$
& $0.5765 \pm 0.0122$ \\

\midrule
\multicolumn{5}{@{}l}{\textit{(c) EchoNet-Dynamic}} \\
\midrule
\textbf{Method}
& \textbf{IoU $\uparrow$}
& \textbf{HD95 $\downarrow$}
& \textbf{ASD $\downarrow$}
& \textbf{BF1 $\uparrow$} \\
\midrule

Joint Training (Offline)
& $0.8240 \pm 0.0052$
& $15.7695 \pm 0.7094$
& $4.4416 \pm 0.1703$
& $0.4996 \pm 0.0036$ \\

Plain Replay
& $0.8313 \pm 0.0037$
& $16.0599 \pm 0.4975$
& $4.4070 \pm 0.1029$
& $0.5073 \pm 0.0039$ \\

Boundary-Balanced Replay
& $0.8270 \pm 0.0029$
& $15.9071 \pm 0.1396$
& $4.4593 \pm 0.0114$
& $0.4908 \pm 0.0016$ \\

EWC
& $0.8323 \pm 0.0076$
& $15.9725 \pm 0.5342$
& $4.3769 \pm 0.1230$
& $\mathbf{0.5122 \pm 0.0075}$ \\

LwF
& $0.3076 \pm 0.0218$
& $44.3878 \pm 2.7851$
& $15.2501 \pm 0.5572$
& $0.1662 \pm 0.0269$ \\

iCaRL-style Replay
& $0.8288 \pm 0.0037$
& $15.7981 \pm 0.1585$
& $4.4135 \pm 0.0349$
& $0.4978 \pm 0.0016$ \\

Weighted-Sampler Replay
& $0.8316 \pm 0.0023$
& $15.6067 \pm 0.7678$
& $4.3140 \pm 0.1394$
& $0.5087 \pm 0.0032$ \\

DER
& $0.8213 \pm 0.0119$
& $41.9954 \pm 44.3794$
& $9.3054 \pm 8.2795$
& $0.4606 \pm 0.0626$ \\

MIR
& $0.8311 \pm 0.0023$
& $15.8751 \pm 0.3035$
& $4.3957 \pm 0.0688$
& $0.5060 \pm 0.0008$ \\

\midrule

\textbf{RAR-Net (Ours)}
 & $\mathbf{0.8331 \pm 0.0026}$
& $\mathbf{15.3991 \pm 0.0927}$
& $\mathbf{4.2779 \pm 0.0449}$
& $0.5094 \pm 0.0042$\\

\bottomrule
\end{tabularx}
\end{table*}

\subsection{Computational Analysis}
\label{sec:computational_analysis}

In addition to segmentation accuracy and CL performance, we examined the computational characteristics of RAR-Net. This analysis is useful because replay-based CL inevitably introduces additional training operations through historical-sample rehearsal. The relevant question is therefore not whether replay is computationally free, but whether the resulting overhead remains practical and whether the proposed method requires architectural expansion as new tasks are encountered. Table \ref{tab:rarnet_computation} summarizes the task-wise training time, peak GPU-memory consumption, and model size of RAR-Net.




\subsubsection{Training Cost}

The computational demand varies substantially across the sequential tasks. CAMUS requires only $786.95\pm10.02$~s of training, whereas CardiacNet is the most expensive stage at $16802.10\pm65.44$~s. EchoNet-Dynamic requires $6765.76\pm16.07$~s. Across the full CAMUS $\rightarrow$ CardiacNet $\rightarrow$ EchoNet-Dynamic sequence, the accumulated training time is approximately $2.44\times10^{4}$~s, corresponding to about 6.77~h.

The difference between tasks primarily reflects the amount and organization of training data processed at each stage rather than a change in network complexity. In particular, RAR-Net uses the same backbone throughout the entire sequence. The relatively small standard deviations in task-wise training time also indicate that the computational requirements remain stable across the three deterministic seeds.

Replay inevitably increases the amount of computation performed during continual adaptation because each update can include both current-task and historical samples. In our setting, the current-task mini-batch size was 16, while the replay mini-batch size was 8. A fixed memory capacity of 1,800 samples was used. Thus, the computational cost of RAR-Net arises primarily from controlled historical rehearsal rather than from expanding the
segmentation network.

\subsubsection{{Evaluation Runtime}}

The mean evaluation time per image was approximately $0.0370$~s for CAMUS, $0.3250$~s for CardiacNet, and $0.0150$~s for EchoNet-Dynamic. These values correspond to approximately 37, 325, and 15~ms per evaluated image, respectively. The recorded values represent the complete evaluation procedure rather than an isolated forward pass. They therefore include data loading, device transfer, model prediction, and metric computation. Consequently, the larger CardiacNet value should not be interpreted as pure neural-network inference latency.

This distinction matters when assessing deployment feasibility. RAR-Net does not introduce a separate inference network, task-specific decoder, or generative module. Once training is complete, prediction is performed using the same fixed segmentation network regardless of the task from which the input originates. The replay mechanisms are used only during training and do not require replay-memory access during inference.

\subsubsection{Memory Requirements}

Peak GPU-memory consumption is 1.6620~GB during CAMUS training, increases to 2.3114~GB for CardiacNet, and remains at 2.2710~GB for EchoNet-Dynamic. The maximum observed GPU requirement is therefore approximately 2.31~GB. The increase after the first task is expected because replay-based stages process historical samples together with current-task data. RAR-Net additionally maintains a bounded replay memory of 1,800 samples. The predefined relative task-allocation weights are 0.25, 0.50, and 0.25 for CAMUS, CardiacNet, and EchoNet-Dynamic, respectively. These weights are renormalized over the tasks observed at each stage. After all three tasks have been observed, the protected quotas are 450 CAMUS samples, 900 CardiacNet samples, and 450 EchoNet-Dynamic samples. The key point is that this storage budget is fixed; it does not grow unbounded as additional stages are learned.

\subsection{Qualitative Results}
\label{sec:qualitative_results}

\begin{figure}[!t]
\centering
\includegraphics[width=0.8\textwidth]{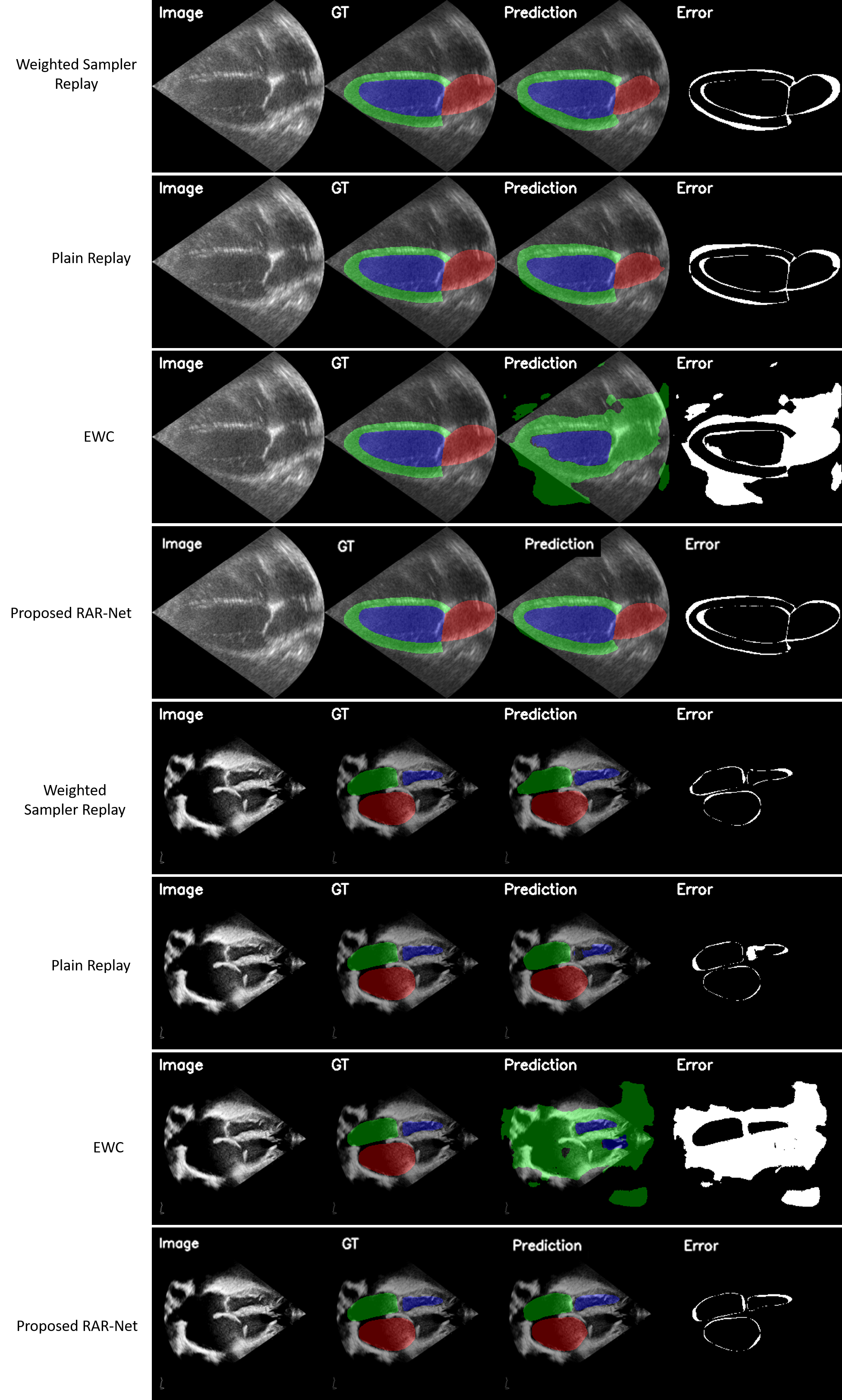}
\caption{Qualitative segmentation comparison on CAMUS (first four rows) and CardiacNet (last four rows). Each row represents a CL method, with columns showing the input image, GT, prediction, and pixel-wise error map. White regions indicate segmentation disagreement with GT.}
\label{quantitative_multistructure}
\end{figure}

\begin{figure}[!t]
\centering
\includegraphics[width=\textwidth]{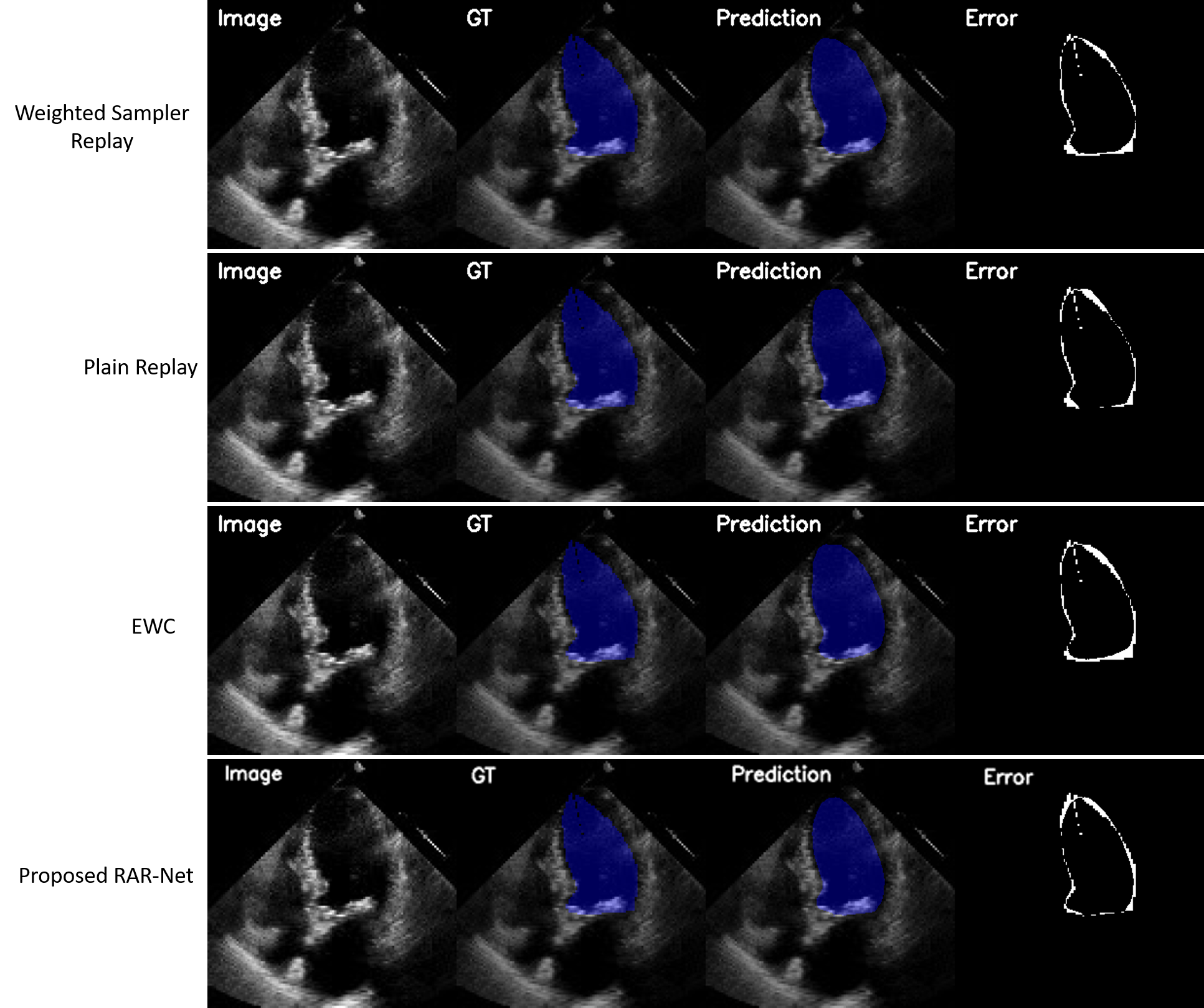}
\caption{Qualitative LV segmentation comparison on EchoNet-Dynamic. Each row corresponds to the CL method indicated in the figure. From left to right, the columns show the input echocardiogram, GT LV segmentation, predicted LV segmentation, and pixel-wise disagreement map. White regions indicate disagreement between the prediction and GT. The predominantly contour-like errors indicate that the remaining disagreement is concentrated mainly near the LV boundary.}
\label{quantitative_echonet}
\end{figure}

The qualitative results provide spatial evidence for the differences observed in the quantitative evaluation. Figures~\ref{quantitative_multistructure} and \ref{quantitative_echonet} present representative predictions from the multi-structure CAMUS and CardiacNet datasets and the LV-only EchoNet-Dynamic dataset, respectively. Each row identifies the corresponding CL method and shows the input echocardiogram, GT annotation, predicted segmentation, and binary disagreement map. White pixels in the error maps indicate locations where the predicted segmentation differs from the reference annotation. Broad filled regions therefore indicate substantial structural disagreement, whereas thin contour-like regions primarily represent localized boundary displacement.

\paragraph{Multi-structure segmentation.}
A clear pattern in Figure~\ref{quantitative_multistructure} is that the differences among the CL strategies are expressed not only through the amount of segmentation error but, more importantly, through its spatial organization. Several methods recover the approximate locations of the cardiac structures, but the weaker predictions occasionally exhibit severe structural failures. In the CAMUS example, the most prominent failure is substantial expansion of the predicted myocardial region into surrounding tissue, producing a broad filled disagreement region. A similar failure is visible in the CardiacNet example, where extensive foreground expansion disrupts the expected multi-structure configuration. Such errors are more severe than small contour offsets because they alter the geometry and spatial organization of the predicted cardiac anatomy.

The stronger replay-based predictions substantially reduce these failures. In particular, Weighted-Sampler Replay preserves much of the global cardiac configuration and demonstrates that appropriately revisiting historical examples can alleviate forgetting. Nevertheless, weighted sampling alone does not explicitly guarantee preservation of the historical task composition or provide structure-specific protection when particular anatomical annotations disappear from subsequent tasks. The qualitative comparison therefore highlights an important distinction between improving replay sampling and controlling the information that is retained and reinforced throughout sequential learning.

RAR-Net exhibits a consistently coherent multi-structure configuration in the displayed examples. The LV, MYO, and LA remain spatially separated, their relative arrangement is preserved, and the predicted regions closely follow the corresponding GT anatomy. Consequently, its disagreement maps are dominated by localized contour differences rather than the broad interior or exterior error regions associated with severe structural failure. The important qualitative pattern is therefore not the complete elimination of segmentation error, but a transition from spatially distributed anatomical errors toward predominantly boundary-localized disagreement.

Another notable pattern concerns the relative vulnerability of the cardiac structures. The LV is generally more stable across the compared methods, whereas MYO and LA are more susceptible to distortion, expansion, or incomplete preservation. Even in predictions exhibiting substantial degradation, a recognizable LV can remain while the surrounding myocardial region becomes excessively thick or spatially extended. This behavior is particularly relevant to the heterogeneous supervision considered in this study. LV supervision remains available in EchoNet-Dynamic, whereas MYO and LA annotations are unavailable in the final task. Consequently, preservation of MYO and LA after learning the LV-only domain provides a more demanding visual indication of retained anatomical knowledge.

The observed behavior is consistent with the complementary mechanisms of RAR-Net. Reliability-aware exemplar prioritization evaluates historical samples using class, boundary, image, anatomical, and temporal information, rather than relying only on their sampling frequency. Protected task-aware memory limits the displacement of historical-domain information, while the forgetting-conditioned controller increases the contribution of replay when degradation on previous tasks becomes more pronounced. Most importantly for the multi-structure examples, structure-aware replay explicitly reinforces MYO and LA using fully annotated historical samples when these structures are no longer supervised by the incoming LV-only domain. Boundary supervision provides an additional constraint on spatial delineation. The interaction of these mechanisms provides a plausible explanation for the observed reduction in large anatomically destructive errors.

\paragraph{EchoNet-Dynamic.}
A different pattern is observed in Figure~\ref{quantitative_echonet}. Because EchoNet-Dynamic provides direct LV supervision and evaluation is restricted to the LV, the qualitative differences among the compared methods are much less pronounced than for CAMUS and CardiacNet. All displayed methods correctly localize the LV and recover its overall ventricular geometry. Their disagreement maps are predominantly thin and peripheral, indicating that the remaining errors arise mainly from local contour displacement rather than complete failure to identify the ventricular region.

The residual disagreement is concentrated particularly near the ventricular boundary, including the lateral wall and apical region, where ultrasound interfaces can be weak or ambiguous. RAR-Net likewise produces a well-localized LV whose overall shape closely corresponds to the GT. Its advantage on this dataset is therefore considerably more subtle than in the multi-structure examples. This is expected because EchoNet-Dynamic is the final task in the primary sequence and directly provides LV supervision during training. Consequently, the compared methods have direct access to the target structure whose segmentation is visualized.

This contrast clarifies the principal role of RAR-Net. The challenge is not only to learn the directly supervised LV in the incoming EchoNet-Dynamic domain, but also to do so without destroying the richer LV-MYO-LA knowledge acquired from the preceding domains. The CAMUS and CardiacNet examples therefore provide more informative visual evidence of historical anatomical retention, whereas the EchoNet-Dynamic examples indicate that this additional stability does not prevent effective adaptation to the newly introduced LV-specific domain.

\paragraph{Comparison with Weighted-Sampler Replay.}
The qualitative comparison with Weighted-Sampler Replay is particularly informative because both approaches use historical exemplars, but differ in how replay information is controlled. Weighted-Sampler Replay can improve historical retention by changing the probability with which stored examples are revisited. The displayed results confirm that such replay can preserve substantial anatomical information. However, sample weighting by itself does not explicitly determine whether individual historical domains remain adequately represented, whether replay strength should increase when measured forgetting becomes larger, or whether structures that lose direct supervision require additional protection.
 
RAR-Net addresses these complementary aspects through protected task-aware memory, forgetting-conditioned replay weighting, and annotation-aware structure preservation in addition to reliability-based exemplar prioritization. The visual consequence in the displayed examples is greater consistency of the retained multi-structure anatomy, particularly for MYO and LA, together with fewer catastrophic foreground expansions. Thus, the qualitative advantage over Weighted-Sampler Replay should be interpreted as more controlled preservation of historical anatomical information rather than simply more frequent replay of selected examples.

\paragraph{Error-pattern interpretation.}
The disagreement maps further distinguish two qualitatively different forms of segmentation error. Broad filled white regions indicate substantial structural disagreement, including severe over-segmentation, under-segmentation, or distortion of an anatomical region. In contrast, thin contour-like regions indicate that the global anatomy has been recovered and that the remaining disagreement is predominantly associated with boundary localization. This distinction complements region-overlap measures because predictions with broadly similar foreground overlap may differ considerably in whether their errors preserve or disrupt anatomical organization.

Across the displayed multi-structure examples, the stronger methods shift the error pattern from large spatially distributed disagreement toward localized contour errors. RAR-Net demonstrates this pattern consistently in the illustrated CAMUS and CardiacNet cases. On EchoNet-Dynamic, where all methods already preserve the directly supervised LV reasonably well, the error maps are predominantly boundary-localized across methods. These observations also support the use of surface- and boundary-sensitive measures such as HD95, ASD, and BF1 alongside Dice-based evaluation.

Overall, the qualitative findings reinforce the quantitative results while clarifying the nature of the improvement. RAR-Net does not eliminate segmentation error, and residual contour discrepancies remain visible, particularly at weak-contrast ultrasound boundaries. Rather, its principal qualitative advantage is the more consistent preservation of the identity, relative position, and geometry of previously learned cardiac structures, together with a reduction in severe spatially distributed failures. This behavior is consistent with the combined action of reliability-aware exemplar selection, protected task-aware memory, forgetting-conditioned adaptive replay, boundary supervision, and annotation-aware structure preservation under sequential domain and supervision shifts.

\subsection{Ablation Study}
\label{sec:ablation}

We investigate the contributions of RAR-Net's principal components and individual reliability factors through two complementary ablation studies. The first examines protected task-aware memory, forgetting-conditioned adaptive replay, and structure-aware supervision, including separate MYO-only and LA-only configurations. The second removes each reliability factor individually while renormalizing the remaining weights. All experiments follow the CAMUS $\rightarrow$ CardiacNet $\rightarrow$ EchoNet-Dynamic sequence and use three random seeds (42, 123, and 2025). The component configurations are summarized in Table~\ref{tab:ablation_configuration}, with quantitative results reported in Tables~\ref{tab:ablation_results} and~\ref{tab:factorwise_reliability_ablation}.

\begin{table*}[t]
\centering
\caption{Configuration of the RAR-Net ablation variants.}
\label{tab:ablation_configuration}

\small
\setlength{\tabcolsep}{6pt}
\renewcommand{\arraystretch}{1.25}

\begin{tabularx}{\textwidth}{
@{}
>{\raggedright\arraybackslash}X
>{\centering\arraybackslash}X
>{\centering\arraybackslash}X
>{\centering\arraybackslash}X
@{}
}
\toprule

\textbf{Variant}
&
\textbf{Protected Memory}
&
\textbf{Adaptive Replay}
&
\textbf{Structure-Aware Replay}
\\

\midrule

A0 Full RAR-Net
& $\checkmark$
& $\checkmark$
& MYO + LA \\

A1 w/o Protected Memory
& $\times$
& $\checkmark$
& MYO + LA \\

A2 w/o Adaptive Replay
& $\checkmark$
& $\times$
& MYO + LA \\

A3 w/o Structure-Aware Replay
& $\checkmark$
& $\checkmark$
& $\times$ \\

A4 MYO-only Replay
& $\checkmark$
& $\checkmark$
& MYO only \\

A5 LA-only Replay
& $\checkmark$
& $\checkmark$
& LA only \\

\bottomrule
\end{tabularx}

\end{table*}

\begin{table*}[!t]
\centering
\caption{Ablation study of the proposed RAR-Net using three
deterministic random seeds. Results are reported as
mean $\pm$ standard deviation.}
\label{tab:ablation_results}

\footnotesize
\setlength{\tabcolsep}{3pt}
\renewcommand{\arraystretch}{1.12}

\begin{tabularx}{\textwidth}{@{}
>{\raggedright\arraybackslash}p{3.5cm}
*{4}{>{\centering\arraybackslash}X}
@{}}
\toprule
\multicolumn{5}{@{}l}{\textit{(a) Continual-learning metrics}} \\
\midrule
\textbf{Variant}
& \textbf{FAA $\uparrow$}
& \textbf{AF $\downarrow$}
& \textbf{BWT $\uparrow$}
& \textbf{FWT $\uparrow$} \\
\midrule

A0 Full RAR-Net
& $\mathbf{0.8718 \pm 0.0029}$
& $\mathbf{-0.0489 \pm 0.0116}$
& $\mathbf{0.0489 \pm 0.0116}$
& $\mathbf{0.4366 \pm 0.0357}$ \\

A1 w/o Protected Memory
& $0.8555 \pm 0.0017$
& $-0.0301 \pm 0.0106$
& $0.0301 \pm 0.0106$
& $\mathbf{0.4366 \pm 0.0357}$ \\

A2 w/o Adaptive Replay
& $0.8689 \pm 0.0015$
& $-0.0407 \pm 0.0167$
& $0.0407 \pm 0.0167$
& $0.4319 \pm 0.0237$ \\

A3 w/o Structure-Aware Replay
& $0.8673 \pm 0.0030$
& $0.0369 \pm 0.0001$
& $-0.0369 \pm 0.0001$
& $0.4247 \pm 0.0107$ \\

A4 MYO-only Replay
& $0.8695 \pm 0.0025$
& $-0.0430 \pm 0.0138$
& $0.0430 \pm 0.0138$
& $0.4296 \pm 0.0227$ \\

A5 LA-only Replay
& $0.8684 \pm 0.0017$
& $-0.0444 \pm 0.0199$
& $0.0444 \pm 0.0199$
& $0.4329 \pm 0.0087$ \\

\bottomrule
\end{tabularx}

\vspace{8pt}

\begin{tabularx}{\textwidth}{@{}
>{\raggedright\arraybackslash}p{3.5cm}
*{3}{>{\centering\arraybackslash}X}
@{}}
\toprule
\multicolumn{4}{@{}l}{\textit{(b) Final Dice after sequential learning}} \\
\midrule
\textbf{Variant}
& \textbf{CAMUS $\uparrow$}
& \textbf{CardiacNet $\uparrow$}
& \textbf{EchoNet-Dynamic $\uparrow$} \\
\midrule

A0 Full RAR-Net
& $\mathbf{0.8977 \pm 0.0024}$
& $\mathbf{0.8096 \pm 0.0055}$
& $\mathbf{0.9081 \pm 0.0023}$ \\

A1 w/o Protected Memory
& $0.8950 \pm 0.0027$
& $0.7670 \pm 0.0060$
& $0.9047 \pm 0.0024$ \\

A2 w/o Adaptive Replay
& $0.8915 \pm 0.0052$
& $0.8073 \pm 0.0076$
& $0.9079 \pm 0.0022$ \\

A3 w/o Structure-Aware Replay
& $0.8941 \pm 0.0008$
& $0.8003 \pm 0.0098$
& $0.9076 \pm 0.0007$ \\

A4 MYO-only Replay
& $0.8946 \pm 0.0026$
& $0.8066 \pm 0.0058$
& $0.9074 \pm 0.0017$ \\

A5 LA-only Replay
& $0.8938 \pm 0.0023$
& $0.8044 \pm 0.0065$
& $0.9072 \pm 0.0018$ \\

\bottomrule
\end{tabularx}
\end{table*}

The component-wise results identify protected memory as the strongest contributor to final segmentation accuracy. Removing this mechanism (A1) reduces FAA from $0.8718$ to $0.8555$, with the largest deterioration occurring on CardiacNet, where Dice decreases from $0.8096$ to $0.7670$. This suggests that task-aware memory allocation helps preserve intermediate-task knowledge that might otherwise be displaced during subsequent training. Disabling adaptive replay (A2) produces a smaller FAA reduction to $0.8689$, accompanied by weaker AF and BWT. The effect is modest, however, indicating that a fixed replay coefficient already retains much of the full model's performance under the evaluated sequence.

\begin{table*}[!t]
\centering
\caption{Factor-wise reliability ablation of RAR-Net.
Results are reported as mean $\pm$ standard deviation over
three deterministic random seeds (42, 123, and 2025).
Each variant removes one reliability factor while
renormalizing the remaining factor weights.}
\label{tab:factorwise_reliability_ablation}

\footnotesize
\setlength{\tabcolsep}{3pt}
\renewcommand{\arraystretch}{1.12}

\begin{tabularx}{\textwidth}{@{}
>{\raggedright\arraybackslash}p{3.5cm}
*{4}{>{\centering\arraybackslash}X}
@{}}
\toprule
\multicolumn{5}{@{}l}{\textit{(a) Continual-learning metrics}} \\
\midrule

\textbf{Variant}
& \textbf{FAA $\uparrow$}
& \textbf{AF $\downarrow$}
& \textbf{BWT $\uparrow$}
& \textbf{FWT $\uparrow$} \\
\midrule

\textbf{Full RAR-Net}
& $\mathbf{0.8718 \pm 0.0029}$
& $\mathbf{-0.0489 \pm 0.0116}$
& $\mathbf{0.0489 \pm 0.0116}$
& $\mathbf{0.4366 \pm 0.0357}$ \\

R1: w/o Class reliability
& $0.8702 \pm 0.0022$
& $-0.0593 \pm 0.0084$
& $0.0593 \pm 0.0084$
& $0.4506 \pm 0.0222$ \\

R2: w/o Boundary reliability
& $0.8701 \pm 0.0031$
& $-0.0504 \pm 0.0041$
& $0.0504 \pm 0.0041$
& $0.4304 \pm 0.0100$ \\

R3: w/o Image reliability
& $0.8640 \pm 0.0033$
& $-0.0381 \pm 0.0169$
& $0.0381 \pm 0.0169$
& $0.4341 \pm 0.0296$ \\

R4: w/o Anatomical reliability
& $0.8696 \pm 0.0024$
& $-0.0436 \pm 0.0072$
& $0.0436 \pm 0.0072$
& $0.4393 \pm 0.0106$ \\

R5: w/o Temporal reliability
& $0.8702 \pm 0.0027$
& $-0.0476 \pm 0.0140$
& $0.0476 \pm 0.0140$
& $0.4334 \pm 0.0309$ \\

\bottomrule
\end{tabularx}

\vspace{8pt}

\begin{tabularx}{\textwidth}{@{}
>{\raggedright\arraybackslash}p{3.5cm}
*{3}{>{\centering\arraybackslash}X}
@{}}
\toprule
\multicolumn{4}{@{}l}{\textit{(b) Final Dice after sequential learning}} \\
\midrule

\textbf{Variant}
& \textbf{CAMUS $\uparrow$}
& \textbf{CardiacNet $\uparrow$}
& \textbf{EchoNet-Dynamic $\uparrow$} \\
\midrule

\textbf{Full RAR-Net}
& $0.8977 \pm 0.0024$
& $\mathbf{0.8096 \pm 0.0055}$
& $\mathbf{0.9081 \pm 0.0023}$ \\

R1: w/o Class reliability
& $0.8956 \pm 0.0025$
& $0.8088 \pm 0.0054$
& $0.9062 \pm 0.0007$ \\

R2: w/o Boundary reliability
& $0.8959 \pm 0.0025$
& $0.8071 \pm 0.0083$
& $0.9073 \pm 0.0012$ \\

R3: w/o Image reliability
& $0.8944 \pm 0.0024$
& $0.7904 \pm 0.0109$
& $0.9074 \pm 0.0002$ \\

R4: w/o Anatomical reliability
& $0.8963 \pm 0.0015$
& $0.8045 \pm 0.0085$
& $0.9079 \pm 0.0023$ \\

R5: w/o Temporal reliability
& $0.8959 \pm 0.0016$
& $0.8057 \pm 0.0083$
& $\mathbf{0.9088 \pm 0.0006}$ \\

\bottomrule
\end{tabularx}
\end{table*}

Structure-aware replay has a more pronounced effect on historical retention than on final segmentation accuracy. Removing it (A3) reduces FAA to $0.8673$ and changes AF from $-0.0489$ to $0.0369$, while BWT decreases from $0.0489$ to $-0.0369$. This reversal suggests that explicit structural supervision contributes to retaining previously learned knowledge when subsequent tasks provide incomplete anatomical annotations. The single-structure variants provide further context. MYO-only replay (A4) achieves slightly higher FAA than LA-only replay (A5), whereas LA-only replay yields marginally more favorable AF and BWT. Neither variant matches the complete model across the reported metrics, suggesting complementary contributions from MYO and LA supervision, although the differences are relatively small.

The factor-wise analysis in Table~\ref{tab:factorwise_reliability_ablation} reveals a less uniform pattern. Removing image reliability (R3) produces the largest reduction in FAA, from $0.8718$ to $0.8640$, alongside a decrease in CardiacNet Dice from $0.8096$ to $0.7904$. This suggests that image-derived cues, including contrast, sharpness, and foreground visibility, are particularly useful for prioritizing historical exemplars under changing ultrasound appearances. Removing anatomical reliability (R4) has a smaller effect, reducing FAA to $0.8696$. The effects of the remaining factors are less consistent: excluding class reliability (R1) slightly improves AF and BWT despite reducing FAA, while removing temporal reliability (R5) marginally improves EchoNet-Dynamic Dice. These differences suggest that the reliability factors provide partly overlapping information and that their current fusion weights are not necessarily optimal.

In summary, the ablations show that the three principal mechanisms influence different aspects of CL. Protected memory contributes most clearly to final accuracy, particularly on CardiacNet, whereas structure-aware replay is associated with more favorable historical retention. Adaptive replay provides a smaller additional benefit. Among the reliability factors, image reliability has the largest observed influence on overall performance, while the remaining cues have more modest or metric-dependent effects. The complete model achieves the highest FAA among the evaluated variants, but the mixed factor-wise results also indicate that further calibration of the reliability weights may improve the balance between historical retention and adaptation.

\section{Discussion}
\label{sec:discussion}

The findings of this study suggest that effective continual echocardiographic segmentation depends not only on preserving historical training examples but also on determining which information should be retained and how it should influence subsequent optimization. RAR-Net demonstrates favorable historical knowledge retention while maintaining competitive adaptation to newly introduced data. However, the differences between its principal components and reliability factors indicate that these benefits arise from several interacting mechanisms rather than a uniformly advantageous selection strategy. This distinction is relevant to CL theory, where the stability-plasticity dilemma remains central, and to medical imaging, where changes in acquisition conditions may coincide with differences in annotation completeness.

\subsection{Stability-Plasticity Balance and Historical Knowledge Retention}

The findings are broadly consistent with the stability-plasticity framework underlying CL.  \cite{kirkpatrick2017overcoming} proposed protecting previously important network parameters to reduce catastrophic forgetting, while \cite{li2017learning} explored preserving historical knowledge through distillation. These approaches demonstrate that previously acquired representations can be protected without unrestricted access to historical training data. Nevertheless, their comparatively weak performance in the present experiments suggests that parameter-level constraints and output consistency alone may be insufficient for continual segmentation across heterogeneous echocardiographic datasets. These constraints preserve selected aspects of the previous model without necessarily maintaining the spatially detailed supervision needed for cardiac structures. This interpretation should not be generalized to all regularization or distillation methods, particularly because their effectiveness depends on the implementation, optimization settings, and underlying domain shifts.

The stronger performance of replay-based approaches agrees with the broader observation that revisiting historical experience can reduce interference between sequentially learned tasks \cite{rolnick2019experience}. Yet the present findings also expose a limitation of conventional rehearsal: retaining examples does not guarantee balanced preservation of every historical domain. The intermediate CardiacNet task is especially informative in this respect. Some replay strategies preserve the earliest dataset relatively well while losing substantial knowledge acquired later in the sequence. RAR-Net's protected memory mechanism appears to address this imbalance by reducing competition between historical tasks for limited storage. From a theoretical perspective, this suggests that memory composition is an important part of the stability-plasticity problem, rather than merely an implementation detail of experience replay.

The comparison with offline joint training provides further context. The relatively small performance gap suggests that carefully selected historical supervision can approximate some benefits of simultaneous multi-domain optimization. However, this finding should not be interpreted as demonstrating equivalence between sequential and offline learning. Joint training operates under different data-access assumptions, and the observed proximity may depend on the similarity of the evaluated domains, the memory budget, and the limited length of the learning sequence. Moreover, the modest advantage over strong replay baselines indicates that RAR-Net's contribution lies primarily in improving the allocation and use of historical supervision rather than introducing a fundamentally different learning paradigm.

\subsection{Reliability-Guided Memory and Anatomical Knowledge Preservation}

The component-wise ablations suggest that memory protection and structure-aware supervision address related but distinct sources of forgetting. Protected memory determines which historical examples remain available, whereas structure-aware replay influences which previously learned segmentation targets are reinforced during optimization. This distinction is particularly relevant under heterogeneous supervision. When later training data provide annotations for fewer cardiac structures, the absence of explicit targets for previously learned regions creates an additional source of uncertainty. Historical image-mask pairs can supply supervision that is no longer available in the incoming dataset.

This interpretation is related to the problem of semantic background shift discussed by  \cite{douillard2021plop} in continual semantic segmentation. Their work demonstrated that changes in class supervision can create ambiguity between background pixels and previously learned categories, motivating feature preservation and pseudo-labeling strategies. RAR-Net addresses a related concern through annotated replay rather than relying exclusively on predictions from an earlier model. The distinction matters because historical annotations provide direct supervision for previously learned structures, whereas pseudo-labels and distillation targets may inherit errors from the teacher model. Nevertheless, the two approaches are not directly equivalent: the present study concerns sequential domain adaptation with heterogeneous anatomical labels, rather than the conventional class-incremental setting \cite{douillard2021plop}. The findings therefore extend the relevance of supervision consistency to a different continual-segmentation scenario without establishing superiority over that specific method.

The structural ablations also suggest that preservation of previously learned anatomy cannot be reduced to a single auxiliary objective with interchangeable targets. MYO-only and LA-only supervision exhibit somewhat different effects on overall accuracy and historical retention. These differences may reflect variations in structural appearance, segmentation difficulty, and sensitivity to incomplete supervision. However, the current evidence does not establish a causal anatomical explanation. In particular, the structure-aware Tversky objective reinforces class-specific segmentation accuracy but does not explicitly constrain myocardial thickness, chamber connectivity, or spatial relationships between cardiac structures. Consequently, anatomical preservation in RAR-Net should be interpreted as retention of previously learned structure-specific segmentation performance, not as a guarantee of anatomical plausibility.

The factor-wise reliability experiments introduce a more critical perspective on the proposed exemplar-selection mechanism. Image reliability appears to provide the clearest benefit, particularly for the intermediate dataset, suggesting that image appearance and visibility may influence the usefulness of stored examples under changing acquisition conditions. This observation is plausible in echocardiography, where contrast variation, speckle, acoustic artifacts, and incomplete visualization can affect segmentation quality. However, image quality is not necessarily equivalent to training value. Difficult or atypical examples may contain information that is essential for generalization, even when their apparent reliability is low. A selection strategy that consistently favors visually clear images could therefore underrepresent challenging patients or acquisition conditions.

The mixed contributions of class, boundary, anatomical, and temporal reliability further challenge the assumption that combining more criteria necessarily produces better exemplar selection. Some factor removals improve historical retention or individual-dataset performance, despite slightly reducing overall accuracy. This suggests possible redundancy among reliability cues, imperfect calibration of their fusion weights, or competing objectives within the memory-selection process. An alternative explanation is that certain factors capture properties strongly associated with particular datasets rather than information that generalizes across domains. The present ablations cannot distinguish these possibilities. Accordingly, the five-factor priority function should be regarded as a structured heuristic supported by empirical evidence, rather than a theoretically optimal estimator of exemplar usefulness.

\subsection{Theoretical Implications, Practical Limitations, and Future Directions}

The broader theoretical implication is that continual medical segmentation involves more than the conventional trade-off between retaining old parameters and learning new ones. Historical knowledge must remain accessible at the level of datasets, anatomical structures, and supervision signals. RAR-Net addresses these dimensions through protected memory, adaptive replay, and structure-specific supervision. However, the evidence for these mechanisms is uneven. Memory protection provides a relatively clear contribution, while adaptive replay yields a smaller incremental benefit. The latter finding raises a legitimate question about whether the additional control mechanism is justified by its computational and implementation complexity. A simpler fixed-replay strategy may remain attractive when the operational cost of adaptive monitoring outweighs its modest improvement.

These observations also have practical implications for medical AI governance. Continual model updates could reduce the need for complete retraining when new institutional data become available, but retaining historical patient images introduces privacy, consent, and data-access concerns. Bounded memory reduces storage requirements without eliminating these obligations. In addition, forgetting-conditioned replay depends on access to historical validation information, which may not remain available across institutions. A deployment-oriented framework would therefore require explicit policies for data retention, secure storage, model updating, and independent evaluation of previously supported populations. Such safeguards are especially relevant when improvements on newly acquired data could conceal deterioration in earlier clinical settings.

Several limitations constrain the interpretation of the findings. First, the evaluation covers a small number of echocardiographic datasets and a short sequential learning scenario. Although the datasets differ in appearance and annotation coverage, they do not represent the full diversity of ultrasound equipment, institutions, patient populations, or clinical acquisition protocols. Second, the memory allocations and reliability weights are predefined, leaving uncertainty about their suitability under different storage constraints or dataset characteristics. Third, the limited number of random seeds provides only an initial estimate of experimental variability. The small differences between several competitive methods and ablation variants require stronger statistical evidence before their practical importance can be established.

A further limitation concerns the relationship between segmentation accuracy and clinical utility. Improved overlap and boundary-distance metrics do not necessarily imply more reliable ventricular volume estimation, functional assessment, or clinical decision-making. Errors concentrated near anatomically important regions may have consequences that are not adequately reflected by aggregate segmentation measures. Similarly, the present experiments do not establish whether the proposed method remains calibrated when encountering low-quality images, uncommon anatomical presentations, or acquisition conditions that differ from those represented in memory. These issues are particularly relevant because reliability-guided selection may unintentionally favor well-represented or visually clearer cases.

Future research should therefore address three specific questions. First, can historical memory be allocated dynamically according to observed forgetting, anatomical coverage, and dataset diversity rather than fixed task quotas? Second, can reliability weights be learned or calibrated from actual replay utility, while preventing systematic exclusion of difficult but clinically relevant examples? Third, can structure-aware supervision be extended with explicit geometric or topological constraints and evaluated through clinically meaningful measurements? These investigations should be accompanied by larger multi-institutional evaluations, patient-level uncertainty analyses, and comparisons under matched memory and computational budgets. Privacy-preserving replay alternatives should also be assessed where historical images cannot be retained.

In summary, the findings support the value of coordinating historical memory selection with structure-specific supervision in continual echocardiographic segmentation. The principal contribution of RAR-Net is not that every component independently improves every metric, but that their combination provides a useful balance between historical knowledge retention and adaptation under heterogeneous supervision. Establishing whether this advantage translates into reliable clinical model updates will require broader validation, more principled reliability estimation, and evaluation criteria that extend beyond segmentation accuracy.

\section{Conclusion}
\label{conclusion}

This study proposed RAR-Net, a reliability-aware replay framework for continual echocardiographic segmentation under domain shifts, limited memory, and heterogeneous anatomical supervision. By integrating multi-factor exemplar prioritization, protected task-aware memory, forgetting-conditioned adaptive replay, and structure-aware supervision, RAR-Net aims to balance knowledge retention and adaptation. Experiments on CAMUS, CardiacNet, and EchoNet-Dynamic demonstrated improved overall performance over the evaluated CL baselines, approaching the offline joint-training reference. Ablation studies highlighted the importance of protected memory and structure-aware replay, while revealing smaller and metric-dependent contributions from adaptive replay and individual reliability factors. Despite these encouraging findings, the evaluation remains limited to three datasets, a short learning sequence, and predefined memory allocations and reliability weights. Future work will investigate adaptive memory management, privacy-preserving replay, stronger anatomical constraints, and broader multi-institutional validation using clinically relevant cardiac measurements.

\section*{Declaration of Generative AI and AI-assisted technologies in the writing process}
During the preparation of this work, the authors used ChatGPT to assist with language editing and improve the readability of the manuscript. After using this tool, the authors reviewed and edited the content as needed and take full responsibility for the content of the published article.

\section*{CRediT authorship contribution statement}
\textbf{Zahid Ullah:} Conceptualization, Methodology, Software, Formal analysis, Investigation, Data curation, Writing - original draft, Writing - review \& editing. \textbf{Sieun Choi:} Conceptualization, Methodology, Formal analysis, Writing - original draft, Writing - review \& editing. \textbf{Jihie Kim:} Formal analysis, Investigation, Funding acquisition, Supervision, Project administration.

\section*{\textbf{Declaration of Competing Interests}} The authors declare that they have no known competing financial interests or personal relationships that could have appeared to influence the work reported in this paper.

\section*{Data availability}
All imaging datasets analyzed in this study are publicly accessible. The CAMUS (\url{https://www.creatis.insa-lyon.fr/Challenge/camus/}), CardiacNet (\url{https://github.com/xmed-lab/CardiacNet}), and EchoNet-Dynamic (\url{https://echonet.github.io/dynamic/}) datasets are available on the platform.


\section*{Acknowledgments}
{This work was supported by the Dongguk University Research Fund of 2026.
This research was supported by the MSIT(Ministry of Science and ICT), Korea, under the ITRC(Information Technology Research Center) support program(IITP-2026-RS-2020-II201789), and the Artificial Intelligence Convergence Innovation Human Resources Development(IITP-2026-RS-2023-00254592) supervised by the IITP(Institute for Information \& Communications Technology Planning \& Evaluation).}






\begin{appendices}

\section{Baselines}
\label{sec:baselines}



We compared RAR-Net with eight reference strategies spanning naive sequential adaptation, offline learning, rehearsal-based methods, regularization, and knowledge distillation. Wherever applicable, the same network architecture, data partitions, optimizer, training schedule, replay capacity, and random seeds were used. This is essential for a fair comparison: changing the memory budget or backbone between methods would make it difficult to determine whether an improvement came from the CL strategy itself.


\paragraph{Joint Training.}
Joint training provides simultaneous access to training samples from all three datasets. Unlike the other methods, it is not a strict CL
approach because historical data remain fully available. We nevertheless include it as an offline reference representing the setting in which the
model does not face the same memory-access restriction as continual learners. Its results should therefore be interpreted as a reference rather than as a
directly deployable CL strategy.

\paragraph{Plain Replay.}
Plain Replay \cite{rolnick2019experience} introduces a bounded memory containing samples from previously encountered tasks. Historical samples are replayed while the model learns the
current task, but no explicit anatomical or reliability criterion is used to prioritize informative examples. This baseline isolates the benefit obtained
simply from retaining a small subset of historical data.

\paragraph{{Weighted-Sampler Replay.}}
{Weighted-Sampler Replay} uses non-uniform sampling to favor informative training examples. Class representation and boundary information are incorporated into the weighting strategy, with respective weights of 0.40 and 0.60. This provides a stronger replay reference than uniform rehearsal and helps determine whether sophisticated memory management is necessary beyond relatively simple class- and boundary-aware weighting.

\paragraph{Elastic Weight Consolidation.}
{EWC \cite{kirkpatrick2017overcoming} is a regu\-larization-based method for CL.}
After learning a task, the importance of individual model parameters is
estimated using the Fisher information. During subsequent training, changes
to parameters considered important for earlier tasks are penalized. EWC
therefore attempts to preserve previous knowledge without explicitly storing
historical images.

\paragraph{Learning without Forgetting.}
LwF \cite{li2017learning} uses knowledge distillation to constrain sequential model updates. Before learning a new task, a frozen copy of the previous model serves as the teacher. The current model is optimized for the new segmentation task while being encouraged to retain the output behavior of the teacher. Unlike replay methods, LwF does not require direct storage of previous training samples.

\paragraph{iCaRL-style Replay.}
The iCaRL-style baseline is adapted from the exemplar-based incremental learning framework of Rebuffi \textit{et al.} \cite{rebuffi2017icarl}.
For the present segmentation problem, a bounded exemplar memory is maintained across sequential tasks, with exemplar management designed to avoid excessive
representation of dominant classes. This provides an exemplar-selection baseline against which the reliability-aware memory strategy of RAR-Net can
be compared.

\paragraph{Boundary-Balanced Replay.}
Boundary-Balanced Replay \cite{ullah2026bbr} explicitly favors samples containing informative anatomical boundaries. Replay priority combines class-related information
with boundary complexity, using weights of 0.40 and 0.60, respectively. This baseline is particularly relevant because cardiac segmentation errors
often occur around thin or ambiguous anatomical structures. It therefore tests whether boundary-aware exemplar selection alone is sufficient to control forgetting.

\paragraph{Dark Experience Replay.}
DER \cite{buzzega2020dark} is a rehearsal-based CL method that preserves historical model responses alongside stored training examples. Unlike conventional replay, which relies primarily on GT supervision, DER regularizes the current model by matching its predictions to previously recorded logits. In our segmentation adaptation, historical logits are stored at a reduced spatial resolution and replayed using a pixel-wise logit regression loss. For LV-only annotations, the regression term excludes unannotated anatomical channels to avoid imposing unsupported class-specific constraints. This baseline evaluates whether preserving historical prediction responses can mitigate forgetting without the reliability-guided exemplar selection and structure-aware replay mechanisms of RAR-Net.

\paragraph{Maximally Interfered Retrieval.}
MIR \cite{aljundi2019online} is a replay strategy that prioritizes historical examples expected to experience the greatest interference from an incoming model update. Rather than sampling memory uniformly, MIR estimates the change in loss on candidate exemplars before and after a virtual gradient step and selects those with the largest predicted loss increase. In our implementation, a candidate subset is drawn from bounded replay memory, and interference is estimated using the segmentation objective and a one-step virtual SGD approximation, while actual model optimization uses Adam. This baseline examines whether interference-driven retrieval provides an effective alternative to the reliability-based memory prioritization adopted in RAR-Net.

\paragraph{RAR-Net.}
RAR-Net is the final proposed continual cardiac segmentation model. Unlike plain or boundary-balanced rehearsal, RAR-Net does not treat all historical
samples or all stages of CL equally. It combines reliability-aware replay selection with task-aware memory protection and adaptive replay weighting. The proposed formulation further introduces structure-aware replay, which places additional emphasis on preserving anatomically vulnerable structures, particularly MYO and LA, when fully annotated historical samples are revisited.

The comparison with Plain Replay tests whether selective replay is preferable to generic rehearsal; comparison with {Weighted-Sampler and Boundary-Balanced Replay} tests whether the additional reliability and structural mechanisms provide benefits beyond simpler priority rules; EWC and LwF provide non-rehearsal references; and iCaRL-style Replay evaluates exemplar-based memory management. Together, these baselines allow the contribution of RAR-Net to be examined
against substantially different strategies for controlling catastrophic forgetting.


\section{Model Parameters}
\label{MP}

RAR-Net contains approximately 7.76 million trainable parameters, and this number remains unchanged throughout the CL sequence. Neither the reliability-aware memory mechanism nor the adaptive replay controller introduces additional trainable modules. Likewise, the structure-aware replay objective modifies the optimization process but does not expand the network.

This fixed parameter count distinguishes RAR-Net from parameter-isolation or dynamically expanding CL approaches, where model capacity can increase with the number of tasks. In RAR-Net, additional CL capability is achieved through selective memory management and training-time regularization rather than through progressive architectural growth. Overall, the computational analysis suggests that the proposed method maintains a moderate resource footprint while preserving a fixed deployment model across all three cardiac domains.

RAR-Net introduces additional training computation through historical replay, reliability-based memory management, and
structure-aware supervision while maintaining a fixed segmentation architecture. Let $B$ and $B_r$ denote the
current-task and replay batch sizes, respectively, $N_t$ the number of current-task samples, $E$ the number of
training epochs, and $\mathcal{C}_f$ the per-sample forward-backward computational cost. For tasks $t>1$,
the approximate training complexity is

\begin{equation}
\mathcal{C}_{\mathrm{train}}^{(t)}
=
\mathcal{O}\left(
E N_t
\left(1+\frac{B_r}{B}\right)
\mathcal{C}_f
\right).
\label{eq:training_complexity}
\end{equation}

With $B=16$ and $B_r=8$, replay increases the nominal sample-processing workload by a factor of $1.5$, excluding additional costs from data loading, validation, reliability scoring, and memory updates. The structure-aware loss operates only on fully annotated replay samples and adds $\mathcal{O}(B_fHW)$ pixel-wise operations, where $B_f\leq B_r$. Persistent replay storage is bounded by $\mathcal{O}(MHW)$ for memory capacity $M$, excluding implementation-specific storage overhead. The network contains approximately 7.76 million parameters and does not expand across tasks. Since replay memory and the adaptive controller are not required at inference, the inference complexity remains that of the underlying segmentation network. Empirically, peak GPU memory ranges from 1.6620 to 2.3114 GB across the three tasks. The measured training times are $786.95\pm10.02$ s, $16802.10\pm65.44$ s, and $6765.76\pm16.07$ s for CAMUS, CardiacNet, and EchoNet-Dynamic,
respectively.




\end{appendices}


\bibliography{sn-bibliography}

\end{document}